\RequirePackage{fix-cm}
\documentclass{svjour3}                     
\smartqed  
\usepackage{graphicx}
\usepackage[table,xcdraw]{xcolor}
\usepackage{mathtools} 
\usepackage{amsfonts,amsmath}
\usepackage{amssymb}
\usepackage{lineno,hyperref}
\usepackage{subfig}
\usepackage{float}
\usepackage{booktabs}
\usepackage{multirow}
\usepackage{makecell}
\usepackage[section]{placeins} 
\usepackage[numbers,sort&compress]{natbib}
\usepackage{xcolor} 
\usepackage[section]{placeins} 
\usepackage{enumerate}
\usepackage{caption}

\usepackage{algorithmic}
\usepackage[ruled, linesnumbered, vlined]{algorithm2e} 

\journalname{Applied Intelligence}
\begin{document}

\title{A GA-like Dynamic Probability Method With Mutual Information for Feature Selection
}

\author{Gaoshuai WANG  \and Fabrice Lauri \and Amir HAJJAM EL HASSANI}
\institute{Gaoshuai WANG, Amir HAJJAM EL HASSANI \at
              Nanomédicine, Imagerie, Therapeutique - EA4662, Université Bourgogne Franche-Comté, UTBM, F-90010 Belfort, France\\
              \email{gaoshuai.wang@utbm.fr, amir.hajjam-el-hassani@utbm.fr}           
           \and
           Fabrice Lauri \at
           Connaissance et Intelligence Artificielle Distribuées (CIAD) - UMR 7533, Université Bourgogne Franche-Comté, UTBM, F-90010 Belfort, France\\
             \email{fabrice.lauri@utbm.fr}
}
\date{Received: date / Accepted: date}
\maketitle
\begin{abstract}
Feature selection plays a vital role in promoting the classifier’s performance. However, current methods ineffectively distinguish the complex interaction in the selected features. To further remove these hidden negative interactions, we propose a GA-like dynamic probability (GADP) method with mutual information which has a two-layer structure. The first layer applies the mutual information method to obtain a primary feature subset.
The GA-like dynamic probability algorithm, as the second layer, mines more supportive features based on the former candidate features.
Essentially, the GA-like method is one of the population-based algorithms so its work mechanism is similar to the GA. Different from the popular works which frequently focus on improving GA's operators for enhancing the search ability and lowering the converge time, we boldly abandon GA's operators and employ the dynamic probability that relies on the performance of each chromosome to determine feature selection in the new generation.
The dynamic probability mechanism significantly reduces the parameter number in GA that making it easy to use. As each gene's probability is independent, the chromosome variety in GADP is more notable than in traditional GA, which ensures GADP has a wider search space and selects relevant features more effectively and accurately. To verify our method’s superiority, we evaluate our method under multiple conditions on 15 datasets. The results demonstrate the outperformance of the proposed method. Generally, it has the best accuracy. Further, we also compare the proposed model to the popular heuristic methods like POS, FPA, and WOA. Our model still owns advantages over them.
\end{abstract}
\section{Introduction}
\label{introduction}
Feature selection is a non-negligible pre-processing step when dealing with high-volume or high-dimensional datasets \cite{sorzano2014survey}. It has been extensively used in various domains, such as image process \cite{too2021hyper, sun2020adaptive}, data clustering \cite{purushothaman2020hybridizing, abasi2021improved,zhong2020subspace}, data classification \cite{ peng2010novel, tang2014feature}. By eliminating the irrelevant features and reducing the dataset's feature scale, it can promote the classifier's performance and convergence speed \cite{lin2015feature}. In total, we categorize feature selection methods into three parts, typically, they are filter methods, wrapper methods, and embedded methods \cite{li2017feature}. Filter methods choose a certain number of features based on criteria like correlation, similarity, information gain, etc; Wrapper methods attempt to evaluate all feature combinations with the classifier which needs lots of computing resources; The embedded methods integrate feature selection and classifier learning into a single process.
Filter methods usually are not as computationally expensive as wrapper methods. While filters lack the evaluation of the selected features' performance on a classifier, wrappers contain a classifier as a part of it, which leads to a better result by wrappers than filters \cite{xue2015survey}.

\begin{figure}
 \centering
 \subfloat[KNN\label{fig:softlabel_voc}]{\includegraphics[width=0.3\textwidth]{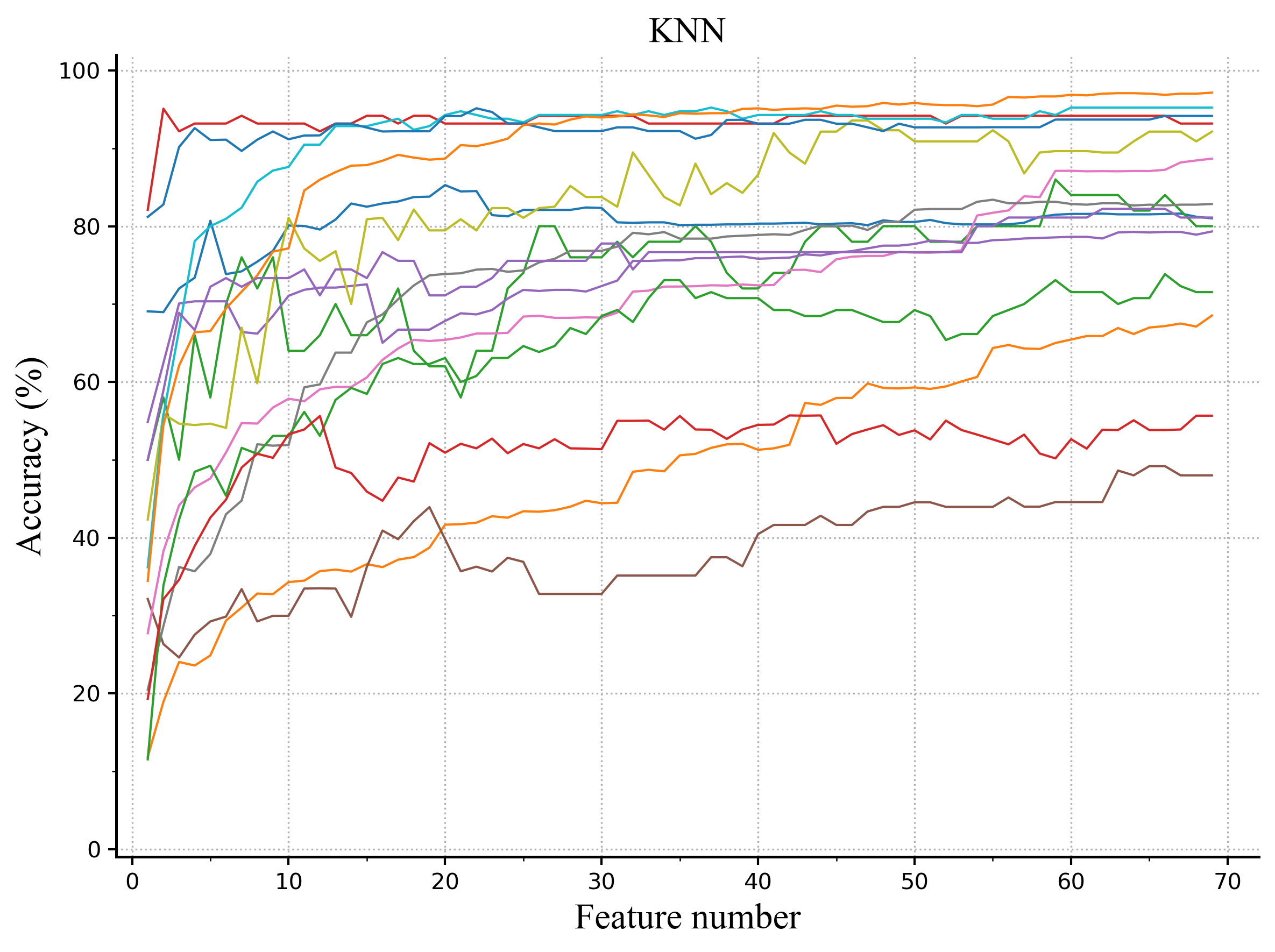}}
 \subfloat[GB\label{fig:softlabel_mit}]{\includegraphics[width=0.3\textwidth]{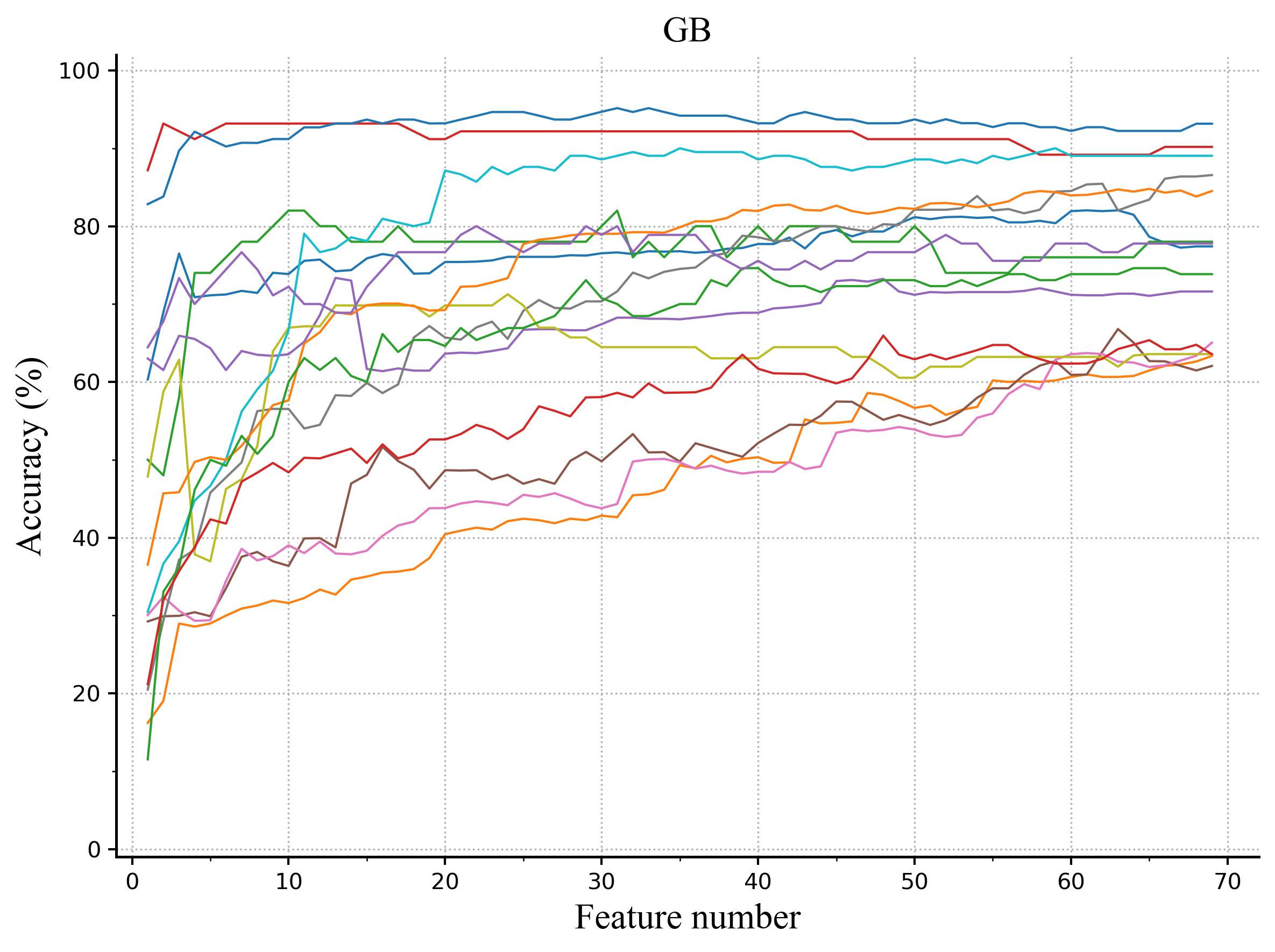}}   
 \subfloat[SVM\label{fig:softlabel_sift}]{\includegraphics[width=0.3\textwidth]{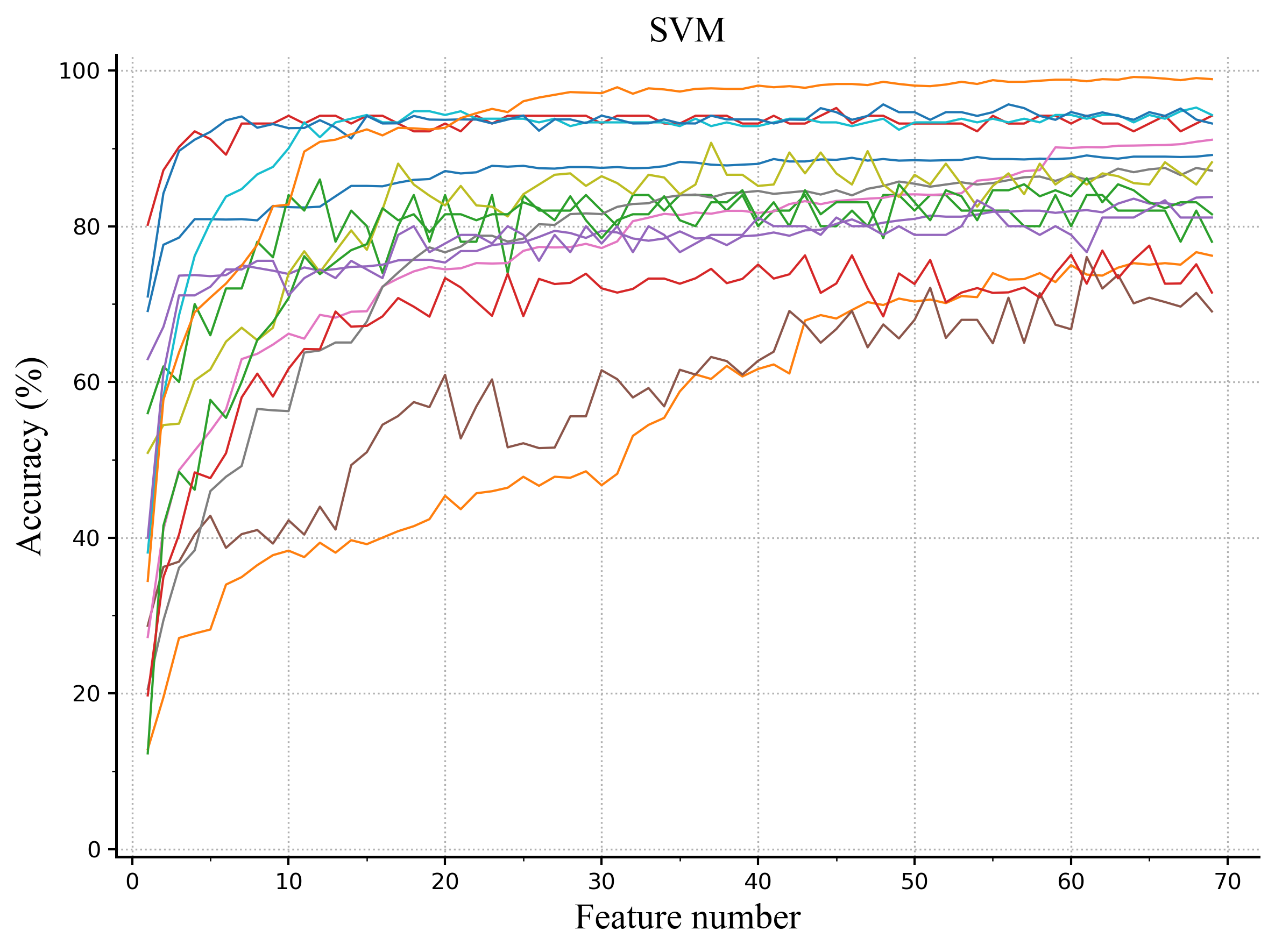}}
 \caption{Variation trend of accuracy with features.}\label{fig1}
\end{figure}

The appearance of Shannon entropy theory \cite{shannon1948mathematical} provides a series of powerful methodologies. Methods like the decision tree and mutual information are designed based on Shannon's theory.
As noted above, the mutual information method should belong to the filter method, which assesses a feature's importance by information gain. If the existence of one feature can tremendously alleviate the uncertainty of the class label, it represents the feature is in favor of the classifier to discriminate the classification.
In recent years, it has gained much attention, and lots of works have demonstrated mutual information's effectiveness \cite{vergara2014review}.
Initially, researchers directly rank the candidate features and adopt the features with top-ranked mutual information as the input of the classifier. However, this method ignored the redundancy existing in features, which will harm the former mutual information selection. Fig.~\ref{fig1} depicts the accuracy growth tendency to the increment of subset features. The accuracy curve's fluctuation means some features are redundant.
Hence, MRMR (minimum redundancy maximum relevance) is proposed by Ding et al. \cite{ding2005minimum} which considers both relevancy and redundancy and achieved huge success. It also inspired researchers to develop more similar methods.

Nevertheless, after summarizing mutual information methods, we found that they only locally release some interactions, for example, these methods often employ the sum of mutual information between the candidate feature and individual feature to replace the mutual information between the candidate feature and the whole of the selected subset. Namely, it's not a perfect replacement and remains some redundancies.

Except for the mutual information methods, researchers also attempt to introduce the heuristic methods such as genetic algorithm (GA) \cite{holland1973genetic}, particle swarm optimization (PSO) \cite{eberhart1995particle} to select optimal features \cite{xue2015survey, kar2016bio}. Heuristic methods are a series of algorithms inspired by creatures and often used in NP-hard problems.
Commonly, the heuristic method owns a strong global search ability which assures they can perform well \cite{chak1995accelerated}. GA is one of the high frequently applied heuristic methods, which is from the biological evolution process \cite{michalewicz1996evolutionary}.
A classical GA comprises several elements, such as chromosome, gene, fitness function, crossover, and mutation. The representative flow generally goes through: initialize the chromosome; select the chromosome by a certain fitness function, generate the new chromosome, crossover, and mutation, repeatedly. The whole process will end until the result converges or reaches the threshold. In most cases, GA method is a feature selector with some classifiers for searching in a global space. However, directly employing GA to choose features still has some defects: it costs too much computation time and hardly makes a balance between local search and global search \cite{el2006hybrid}. Besides, there are plenty of parameters lying in the traditional GA methods which adversely increase the usage difficulty \cite{hansen2015evolution}.

Considering the characteristics of filter and wrapper methods, the hybrid algorithm has increasingly gained attraction from researchers and various combined methods are constantly emerging. Admiring the merits of the hybrid algorithm, we propose a novel GA-like dynamic probability algorithm with mutual information for feature selection.
In our method, the mutual information method is employed to rank all features and then selects top-$N$ features. It can be treated as a global feature selector.
Then, the GA-like method exploits further search locally and offers the selected features to a classifier.
In terms of the GA-like method, it abandons the crossover and mutation operators and generates the new chromosome depending on each gene's probability. It's easier for setting parameters and strengthen our proposed method's local search ability.

In this research, our contributions are:
\begin{enumerate}[1)]
\item We propose a two-layer feature selection model which both considers the global search and local search.
\item To accelerate the feature selection, we employ the MRMR method as the global selector. 
\item We propose a new GA-like dynamic probability method that is simpler compared to GA for searching features locally.
\item Dynamic probability update mechanism replaces the complex operators in GA.
\item Lots of experiments are exploited to demonstrate the characteristics and performance of the proposed method.
\end{enumerate}

The rest of this paper is arranged as follows. 
In Section \ref{Related works}, we present some related works on mutual information and heuristic algorithm in feature selection. 
Section \ref{Preliminaries} presents the main concept of the GA method and other methods. Section \ref{Proposed method} detailedly introduces our proposal GA-like method and the whole process of feature selection. In Section \ref{results}, the datasets and processing method are described. After, we compare our method to the classical GA method with various conditions. Finally, Section \ref{Conclusion} concludes this paper and refers to the possible further work.

\section{Related works}
\label{Related works}

Numerous works have been done since the effectiveness of genetic algorithms and mutual information algorithms. This section will present researchers' efforts at improving and applying them in various domains.

As for the genetic algorithm, researchers often improve its crossover and mutation operators which have a profound effect on its performance.
Xue et al. \cite{xue2021adaptive} proposed an adaptive crossover operator based on a genetic algorithm for feature selection. In their method, they adopted five crossover operators with proper probability and each of them will be selected in different process stages because they consider the single crossover operator is problem-dependent. Their method was compared with
five well-known evolutionary multi-objective algorithms on ten datasets. The results demonstrated that the proposed algorithm can remove redundant features and keep low classification error and is superior to other methods. Manzoni et al. \cite{manzoni2020balanced} used three different balanced crossover operators in cryptography and design theory for the sake of their balancedness constraint. They found that the balanced crossover operators perform better than the one-point crossover. Sun and Liu \cite{sun2019self} considered that premature convergence often happened on classical genetic algorithms due to the rapid multiplication of outstanding individuals. They put up a method to qualify the diversity and similarity between adjacent generations and adjust the crossover and mutation probability dynamically. Their method is proven that greatly improved the convergence speed and global search ability.
In \cite{amini2021two}, GA was used to select features directly for the second elastic net which was flexible in adjusting the penalty terms in the regularization process and time efficiency. Moreover, researchers also explored the initialization \cite{mustafi2019hybrid, gupta2012overview}, encoding/ decoding schemes \cite{kumar2013encoding,costa2010new} and other aspects.

About mutual information, reducing irrelevant features, and keeping supportive features are the core of researchers' work.
Maes et al. \cite{bahl1986maximum} applied the MIM method for estimating the parameters of the hidden Markov model in 1986. It promoted the probability largely when parameters were estimated by
maximum likelihood estimation. However, the feature relevancy ignored by MIM makes mutual information still has a huge improvement space. Battiti \cite{battiti1994using} proposed the mutual information feature selection(MIFS) which considers both the relevancy and redundancy. $\beta$ represents the parameter of the feature redundancy. But, it's difficult to choose a proper value of $\beta$. The same problem also happens on its variants, MIFS-ND and MIFS-U \cite{hoque2014mifs,kwak2002input}.
To balance the relationship between feature relevancy and feature redundancy, Pent et al. \cite{peng2005feature} set the $\beta$ equal to the inverse of the selected feature scale in MRMR.
After that, other methods such as conditional information feature extraction (CIFE) \cite{lin2006conditional}, double input symmetrical relevance (DISR) \cite{meyer2008information}, and Joint mutual information (JMI) \cite{yang1999data} are brought up which emphasised different aspects.

Except for the single genetic algorithm and the single mutual information method, the combined methods of genetic algorithm and mutual information attracted many researchers' attention.
Jadhav et al. \cite{jadhav2018information} carried out the information gain and genetic algorithm in credit rating. The gained information availed reducing the computing complexity and the genetic algorithm further promoted the classifiers' performance. Salesi et al. \cite{salesi2021taga} used a two-layer framework that combined the MRMR method and tabu asexual genetic algorithm to enhance the generalization power. The results show that their algorithm was superior to other conventional and state-of-art algorithms.
Huang et al. \cite{huang2007hybrid} made the genetic algorithm as the first layer and mutual information as the second layer. Their experimental results show that the hybrid method achieved excellent classification accuracy.

\section{Preliminaries}
\label{Preliminaries}

Darwin's theory of evolution by natural selection inspires the appearance of GA that mimics nearly all the stages of gene heredity. Generally, gene heredity is a complex process in biological view, which mainly takes account of several vital parts: such as chromosome representation, crossover, mutation, and fitness selection.
Chromosome representation transfers the specific problem into chromosome format; Crossover indicates the parent chromosomes change their part gene in the same position and produce new chromosomes; Mutation represents some positions of the chromosomes are randomly flipped on a certain probability; Fitness selection will discard or change gene's existence probability in the next off-springs,
which assures the gene's environmental adaptability.

\begin{figure*}[htb]
    \center{\includegraphics[width=0.5\textwidth]
    {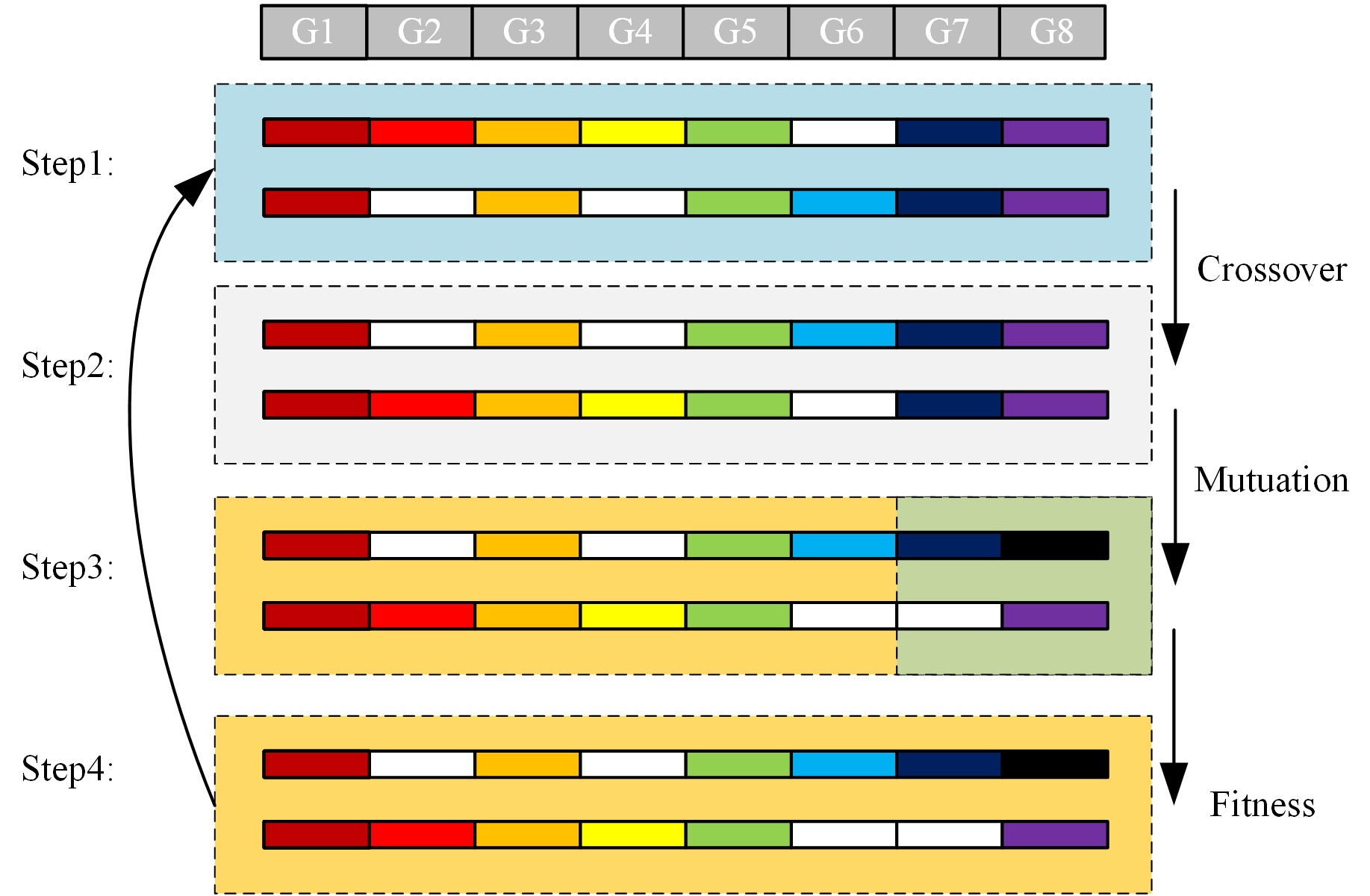}}
    \caption{\label{figure1} Chromosome, gene and operators.}
\end{figure*}

Fig. \ref{figure1} depicts several relative elements and processes in GA. The color bar means a chromosome where several genes exist.

\begin{enumerate}[1)]
\item \textbf{Chromosome representation}:
In the beginning, we need to transform our specific problem into a chromosome which is an essential step because it determines the successful application of GA. Different from the creature which usually contains dozens of chromosomes, in GA, one chromosome represents the whole resolution space, numerous chromosomes will form a population. In terms of feature selection, we can regard each feature as a chromosome gene located in a fixed position. A feature only has two states: be chosen or not, in Fig. \ref{figure1}, color block means the gene is selected. There are lots of chromosomes in each process which assures the global search ability of GA.
\item \textbf{Crossover}:
When the new generation chromosomes are produced, the crossover will happen among the chromosomes, for example, in Fig. \ref{figure1}, Step1 to Step2, two parents' chromosomes generate two children. G1, G4, and G6 are exchanged between the parents. Crossover increasingly enhances solution space's variety and avoids the local optimum. But, it also adds more parameters in the setting process, the crossover proportion, and position.

\item \textbf{Mutation}:
Mutation means a novel gene appearing or some changes beyond the crossover process. In Fig. \ref{figure1}, the mutation process exists in Step2 to Step3. G7 disappears in one chromosome and a new gene appears in the G8 position. It must be emphasized that there is no new gene's appearance in the feature selection problem.

\item \textbf{Fitness selection}:
After the previous processes, fitness selection evaluates each individual's performance. In the natural environment, an excellent gene will strengthen the individual's survival ability, then, its heredity probability increases, correspondingly, and the population gradually owns the adaptability. In GA, when dealing with the fitness result, people usually take two kinds of actions: One is discarding the bad performance individuals directly; Another is reducing their chances of reproduction.
\end{enumerate}
Loop these four steps repeatedly until trigger the terminal conditions.

PSO was brought up by Kennedy and Eberhart. The basic elements named particles are placed in the search space and the fitness function evaluates its current location. Then, each particle decides its velocity and direction in the space relying on its current location and the best location with some members of the swarm. All the particles that have been moved are a complete iteration. Eventually, all the particles like a flock of birds seek food until they find the fitness function‘s optimum.

Flower Pollination Algorithm (FPA) \cite{yang2012flower} was inspired by the natural pollination process.
FPA incorporates biotic cross-pollination and abiotic pollination as the global search and local search processes. Pollen-carrying pollinators such as insects or the wind move in a way that obeys Levy flights. Besides, the switching of local pollination and global pollination is controlled by the probability $p \in [0,1]$. The whole process intimates the plants' pollination.

The whale optimization algorithm (WOA) \cite{mirjalili2016whale} mimics the social behavior of humpback whales. The way that whales hunt their prey is named the bubble-net feeding method. In WOA, each solution is regarded as a whale. The whale moves its place in the search space considered as a reference to the best element of the group. Two search mechanisms are used by the whales to hunt. In the first one, the preys are encircled and the second is the bubble net.

\section{Proposed method}
\label{Proposed method}

Although GA has a puissant ability for global search, it's exhaustive when the scale of the dataset's dimensionality increases to several thousand.
It's not easy for GA to obtain a pretty good result. Considering the high effectiveness of filter methods such as mutual information, we combine these two methods and form a two-layer hybrid method. In mutual information methods proposed with different considerations, we found that MRMR often is one of the baselines and its performance is well.
Thus, in the first layer, we adopt the MRMR method for ranking all the candidate features and select top-$N$ features as the second layer's input. Next, the proposed GA-like method with a specific classifier to search for optimal combinations until it triggers some termination conditions.

Let $X =\{x_1, x_2, ..., x_n\}$ be a feature vector and $p(x_i)$ is the probability of $x_i$, let $Y=\{y_1, y_2, ..., y_n\}$ as the class vector. 
The mutual information between $X$ and $Y$ is calculated in Eq. \ref{eq3}:

\begin{equation}  
I(X,Y)= H(X) - H(X|Y)\label{eq3}
\end{equation}

\begin{algorithm}
    \SetKwData{Left}{left}
    \SetKwData{This}{this}
    \SetKwData{Up}{up} 
    \SetKwFunction{Union}{Union}
    \SetKwFunction{FindCompress}{FindCompress} 
    \SetKwFunction{I}{I} 
    \SetKwInOut{Input}{input}
    \SetKwInOut{Output}{output}   
    \Input{The traning dataset $D$ with original feature set $F=\{f_1,f_2...f_n\}$, class label $Y$ and the required feature number $T = 70$
													} 
    \Output{The selected subset feature $S_T$} 
    \BlankLine 
    
    \emph{$S_T \leftarrow$ $\varnothing$}\; 
	 \lForEach{ $f_i$ in $F$}{ $MI_i =I(f_i;y)$}
		\emph{$f \leftarrow \  max( \ MI)$}\;
		\emph{$S_T \cup f$ }\;
		\emph{$F = F - f$}\;
    \For{$i\leftarrow 2$ \KwTo $T$}
    { 
		
        \For{each $f_i\  in \ F$}
        {
				$J(f_i)$ = $MI_i$ - $\frac{1}{|S_{T}|}\sum_{f_j\in S_{T}}I(f_j; f_i)$\;
			}
        \emph{select \ $f_s$ \ from $J(f_i) \ with \ the \ largest \ value$}\;
			\emph{$S_T = S_T\cup f_s$}\;
			\emph{$F = F - f_s$}\;} 
    \Return $S_T$
    \caption{MRMR \cite{ding2005minimum}}
    \label{alg1} 
\end{algorithm}

\subsection{GA-like dynamic probability method}
GA has already been applied in many areas because it can obtain pretty results in a short time even with a complex task. In GA, the typical processes are crossover, mutation, selection, and reproduction.
The crossover and mutation make the off-springs inherit merits from their parent chromosomes and surpass them in variety.
Meanwhile, there is an inevitable issue that we should have a proper change rate and position during the crossover and mutation processes which easily lead to results unpredictable. Hence, to discover better operators, researchers designed lots of rules in crossover and mutation processes, such as the single point crossover, uniform crossover, cycle crossover; Inversion mutation, the bit flipping mutation, etc. What's more, the selection method is also mandatory when the algorithm produces a new generation. As mentioned above, the GA parameters setting is so complex that researchers need to spend much attention on this part.

Furthermore, we think that GA could not balance the trade-off well locally and globally as there is no feature pre-screening, which will hinder its application.
For simplifying the traditional GA, we propose a GA-like method based on a dynamic probability mechanism. In the whole process, the birth of each new generation is according to their probability.
\begin{enumerate}
\item \textbf{Encode and Decode}. The feature selection problem is simple to encode, we just employ the binary encode to process the feature subset. Its merit is that each feature is independent which assures chromosome variety. In the binary encode, ‘1' represents the feature participating to classify the label; ‘0’ indicates the feature is not selected. Decode is transforming the binary chromosome to the feature subset. We define the $i$ th chromosome in population as $\theta_i$, the $g$ th gene or feature in chromosome $\theta_i$ as $\theta_i^g$. $P_t(\theta_i^g)$ is the probability of $i$ th chromosome's $g$ th gene at $t$ th generation.

\item \textbf{Initialization}. Initialization demonstrates whether a feature should be selected or not in the beginning. In most cases, GA adopts the random initialization method. If the scale of the population is large enough, chromosome variety enhances its global search ability. In our proposed method, whether a feature appears in the chromosome depends on its probability. All the features have the same probability. The too-large probability will degrade the selection function and the too-small probability will decrease the variety of chromosomes.

\begin{equation}
P_0(\theta_i^g) = 0.65   \qquad       0	\leq i<n, 0	\leq g<m
\end{equation}

where $n$ is the number of population, $m$ is the amount of gene or feature in one chromosome. $P_0$ is the probability at $0$ th generation.
\item \textbf{Fitness selection}. In our method, we introduce the Gaussian bayesian (GB), Support vector machine (SVM), and K-nearest neighbors (KNN) as the evaluation function. The higher accuracy means the chromosome has a better performance. In each Fitness selection, 1/3 population will be discarded. The rest of the population will take part in the probability update. Among these populations, The best performance subset will be kept. If a chromosome in the next generation is better than the current chromosome, it will be replaced.

\item \textbf{Probability update}. After the discarding process, relying on remained chromosomes, each gene's probability will be updated. If the existence of one feature promotes accuracy, which indicates the feature is positive in classification, its probability will increase in the next generation. The probability $P_t(\theta_i^g)$ is calculated as:

\begin{equation}
E_{t-1}({\theta_i^g})= 
\begin{cases}
1& \text{gene selected}\\
0& \text{gene not selected}
\end{cases} \\
\end{equation}

\begin{equation}
P_t(\theta_i^g) =  (\sum_{i=1}^{\frac{2n}{3}}E_{t-1}({\theta_i^g})) /\frac{2n}{3}
\label{lable5}
\end{equation}

\item \textbf{Chromosome replication}. Different from the popular chromosome replication method where the new chromosome is produced from a pair of parents. In our proposed method, there is no parent chromosome, all the genes or features only rely on each probability.
\end{enumerate}

\begin{figure}[htb]
    \center{\includegraphics[width=0.5\textwidth]
    {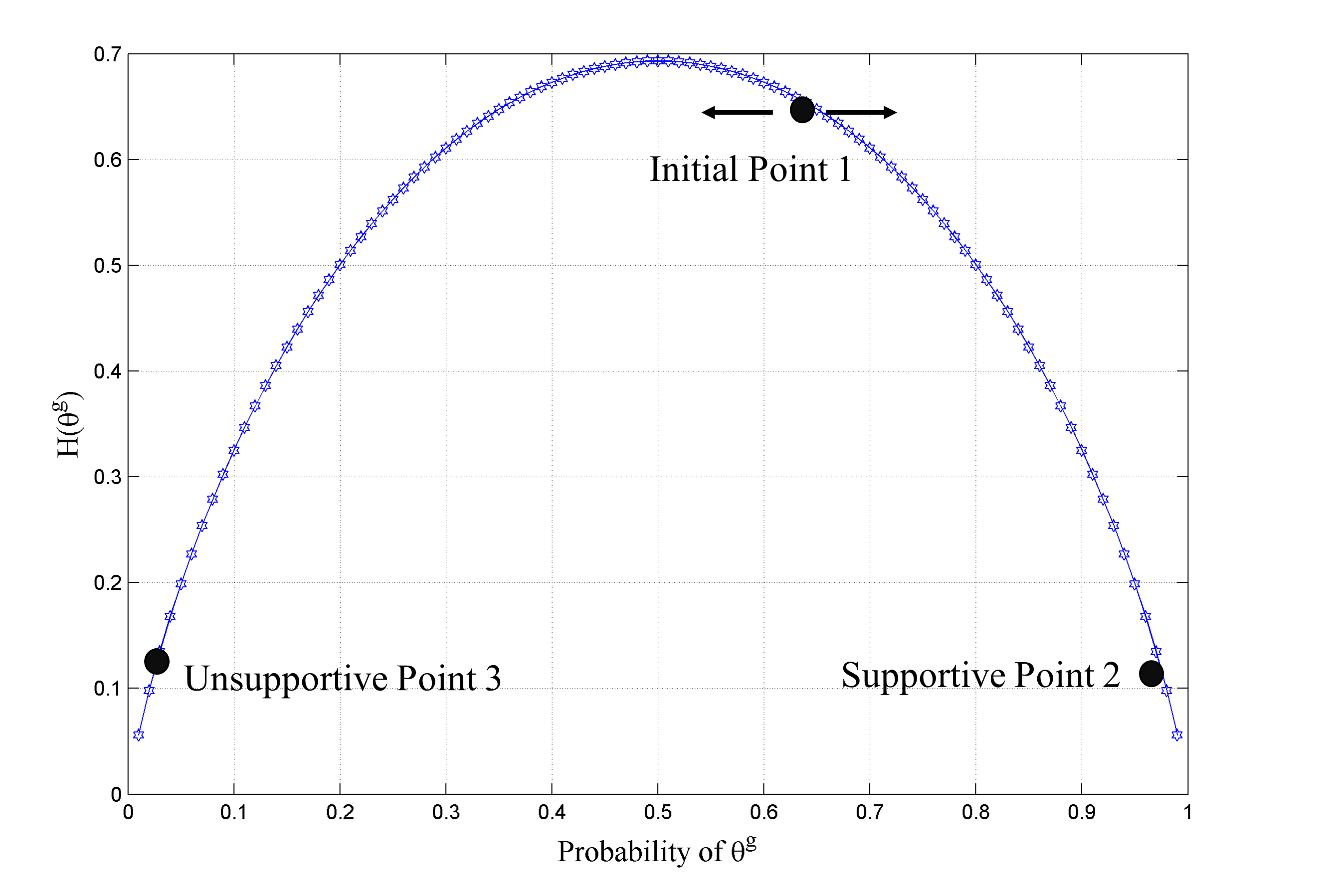}}
    \caption{\label{figure3} The information entropy of gene $g$. Point 1 is the initial information entropy. With the growth of generations, each gene's probability tends to be stable, these supportive genes' information entropy will be close to Point 2 and the redundant genes' information entropy will reach Point 3. }
\end{figure}

To have a better understanding of the variety, we apply the information entropy to calculate the information volume of each gene. At generation $t$, the information entropy of gene $g$ and a chromosome are computed by Eq.~\ref{lable7} and Eq.~\ref{lable8}.
In Fig.~\ref{figure3}, we present one gene's information entropy variation with its probability.

\begin{equation}
H_t(\theta^g) =  -P_t(\theta^g)\ln(P_t(\theta^g))-(1-P_t(\theta^g))\ln(1-P_t(\theta^g))
\label{lable7}
\end{equation}

\begin{equation}
H_t(\theta) =  \sum H_t(\theta^g)  
\label{lable8}
\end{equation}

In contrast to the classical GA, seemingly, their selection method, such as the elite selection or probabilistic selection is similar to our proposed method. Information entropy also can estimate GA's information volume.
However, each gene's probability is not independent on account of the crossover mechanism that decreases off-springs uncertainty and the mutation only increases the uncertainty limitedly.
What's more, the proposed method needn't keep the aforementioned elite generation chromosomes for obtaining good results, all training process influences are in the probability.
Thus, we deem that our proposed method has a more wide search area. We display the Pseudo-code of GADP in Algorithm \ref{alg2}.

\begin{algorithm}
    \caption{Pseudo-code for GADP}
    \label{alg2} 
    \SetKwData{Left}{left}
    \SetKwData{This}{this}
    \SetKwData{Up}{up} 
    \SetKwFunction{Union}{Union}
    \SetKwFunction{FindCompress}{FindCompress} 
    \SetKwFunction{I}{I} 
    \SetKwInOut{Input}{input}
    \SetKwInOut{Output}{output} 
    
    \Input{Feature set $F=\{f_1,f_2...f_n\}$, class label $Y$, population set $N$, initial probability set $P$, iteration number $I$, 
			threshold $t$, discard rate $l$										} 
    \Output{The selected subset feature $S_F$} 

	$N.size \gets 50$, $P.size \gets len(F)$, $I \gets 50$, $t \gets 10$ \;
    Push  $N.size$ empty chromosomes into population set  $N$ \;
   $P \gets \{0.65,0.65...0.65\}$ \;
    $i \gets 0$, $Best_{set}\gets \varnothing $ \;
	\While{$i < I$}{
    \ForEach{chromosome $\theta$ in population set $N$} {   
		\ForEach{gene $g$ in chromosome $\theta$}{
				$g \gets 1$ at the probability of $P(g)$   \tcp{ Produce new chromosome}
             }
         }
    \ForEach{chromosome $\theta$ in population set $N$}{
            $F_{position} \gets Decode(\theta)$\;
			  $x_{train} \gets F[:,F_{position}]$\;
            $E(\theta) \gets EvulationFunction(x_{train},Y)$ \;
         }
      $E(\theta) \gets Sort(E(\theta))$\;
      Discard \quad  $N/3$ \quad chromosomes based on $E(\theta)$\;     
     $Best_{set} \gets$ best performance chromosome \;
    Break the condition If $Best_{set}$ doesn't update more than $t$ times continuously\;
    \ForEach{gene $g$ in one chromosome}
         {
		  $P(g) \gets Eq. ~\ref{lable5}$   \tcp{ Probability update}        
         }
    Remove all chromosomes and push $N.size$ empty chromosomes into population set  $N$ \;
	}
    \Return $Best_{set}$
\end{algorithm}

\begin{figure*}[htb]
    \center{\includegraphics[width=0.8\textwidth]
    {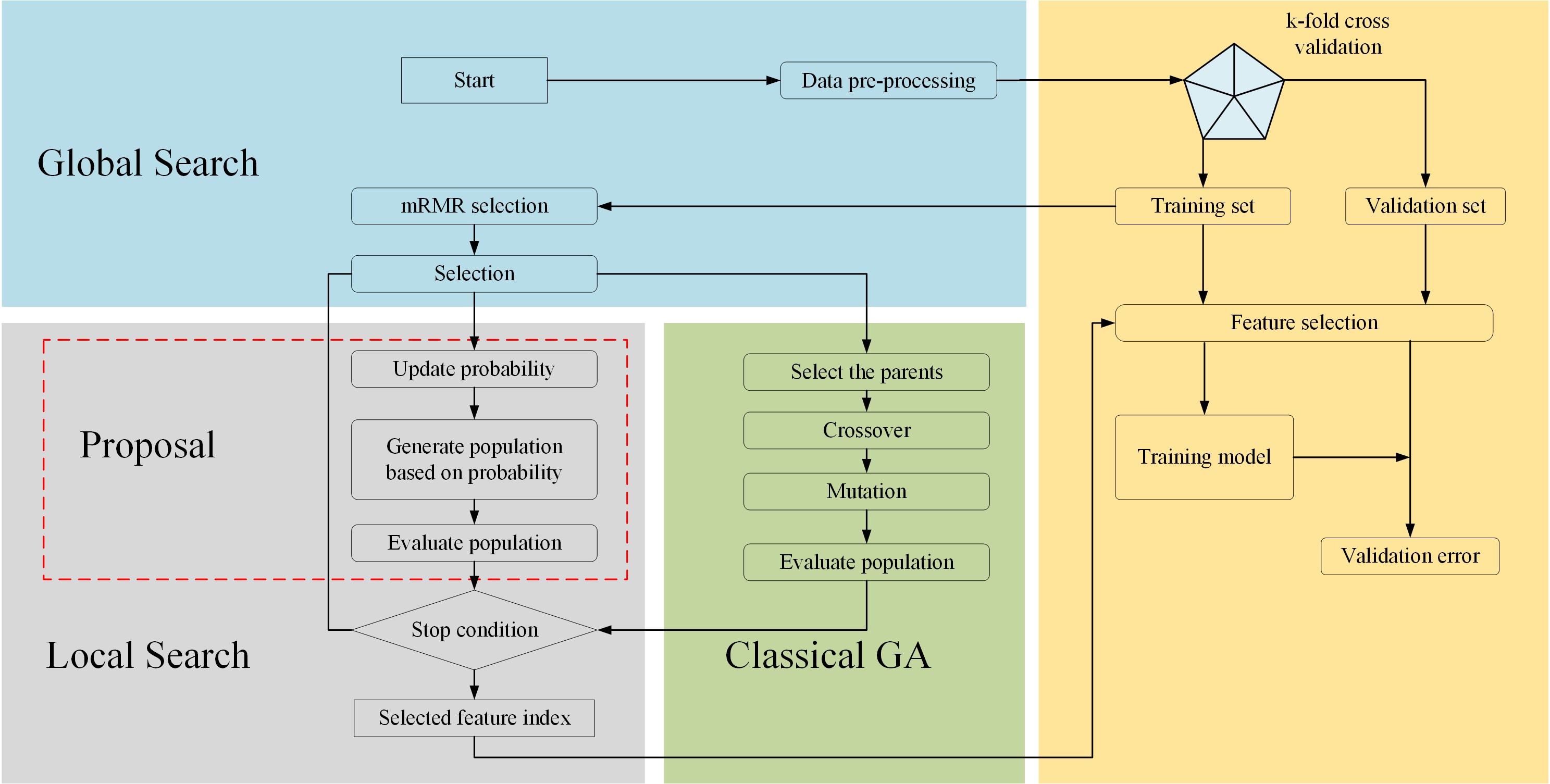}}
    \caption{\label{figure4} The flowchart of proposed method and classical GA}
\end{figure*}

Finally, we present the flowchart of our proposed method and classical GA with mutual information process in Fig. \ref{figure4}. It could be brutally divided into two search parts: global search and local search. MRMR works for the global search part, which reduces the scale of search space largely; GADP is responsible for further removing redundant features. Compared to the classical GA, its structure is more compact.

\section{Experimental results and analysis}
\label{results}
\subsection{Metrics}
We use the accuracy, F1 and AUC as our evaluation indicators, their calculation formulas are as follows:

\begin{equation}
Accuracy =  \frac{TP + TN}{TP + TN + FP + FN}
\end{equation}

\begin{equation}
Precision =  \frac{TP}{TP + FP},  Recall =  \frac{TP}{TP + FN}
\end{equation}

\begin{equation}
F1 =  \frac{2 * Precision * Recall }{Precision + Recall}
\end{equation}

\begin{equation}
AUC = \int_{0}^{1} t_{pr}(f_{pr})d f_{pr} = P(X1 > X0)
\end{equation}

\noindent where $TP, TN, FP$ and $FN$ are the true positive, true negative, false positive and false negative, respectively.
The $t_{pr}$ is the true positive rate, $f_{pr}$ is the false positive rate, and $P(X1 > X0)$ are the probability of the confidence scores $X1 > X0$ for a negative and positive instance.
The AUC value describes the probability that a classifier ranks a randomly chosen positive instance higher than a randomly chosen negative one.
\subsection{Dataset and pre-processing}

To verify the generalization of our proposed method, we employ 15 datasets to assess its performance. We summarize their characteristics of instance, feature, and classes in Table \ref{dataset}.
Three datasets are the binary classification task, and the rest of them are the multi-classification task.
Some datasets are continuous types, which will hinder the accuracy of mutual information selection because continuous types could paralyze mutual information calculation. Therefore, we need to redivide all feature attributes into 10 intervals. The first step is normalizing the attribute as Eq.~\ref{lable9}. Afterward, multiply $x^{*}$ by 10, so all the attributes will fall into the 10 regions. It's a necessary process when using mutual information in a continuous dataset. K-fold cross-validation is used in this research, $K = 5$.

\begin{table*}[htbp]
\caption{The summary of benchmark datasets}
\renewcommand{\arraystretch}{1.0}
\begin{center}
\resizebox{1.0\columnwidth}{!}{
	\begin{tabular}{c c c c c c}
	\hline
		Number 	&	Name	&	Instances	&	Features	&	Classes	&	Notes	\\
    \hline
1	&	PCMAC	&	1943	&	3289	&	2	&	\{1: 982, 2: 961\}	\\
2	&	Isolet	&	1560	&	617	&	26	&	balanced	\\
3	&	GLIOMA	&	50	&	4434	&	4	&	\{4: 15, 1: 14, 3: 14, 2: 7\}	\\
4	&	Prostate\_GE	&	102	&	5966	&	2	&	\{2: 52, 1: 50\}	\\
5	&	RELATHE	 &	1427	&	4322	&	2	&	\{1: 779, 2: 648\}	\\
6	&	TOX\_171	&	171	&	5748	&	4	&	\{1: 45, 2: 45, 4: 42, 3: 39\}	\\
7	&	USPS	&	9298	&	256	&	10	&	\{1: 1553, 2: 1269, 3: 929, 5: 852, 7: 834, 4: 824, 10: 821, 8: 792, 6: 716, 9: 708\}	\\
8	&	CNAE-9	&	1079	&	855	&	9	&	\{7: 120, 6: 120, 5: 120, 4: 120, 8: 120, 3: 120, 2: 120, 9: 120, 1: 119\}	\\
9	&	lung\_discrete	&	73	&	325	&	7	&	\{7: 21, 4: 16, 6: 13, 5: 7, 1: 6, 3: 5, 2: 5\}	\\
10	&	warpPIE10P	&	210	&	2420	&	10	&	balanced	\\
11	&	lung	&	203	&	3312	&	5	&	\{1: 139, 3: 21, 4: 20, 2: 17, 5: 6\}	\\
12	&	COIL20	&	1440	&	1024	&	20	&	balanced	\\
13	&	warpAR10P	&	130	&	2400	&	10	&	balanced	\\
14	&	Yale	&	165	&	1024	&	15	&	balanced	\\
15	&	Brain\_Tumor\_1	&	90	&	5920	&	5	&	\{5: 60, 1: 10, 2: 10, 4: 6, 3: 4\}	\\
\hline
	\end{tabular}
}

\label{dataset}
\end{center}
\end{table*}

\begin{equation}
x^{*} = \frac{x - x_{min}}{ x_{max}- x_{min}}
\label{lable9}
\end{equation}

\subsection{Experimental setting}

Some of our datasets are unbalanced, thus, we use the macro averaging metrics to evaluate the algorithm's performance. In the mutual information process, we only select the first 70 features provided to GADP and other compared algorithms. Essentially, GA belongs to the wrapper method, the classifier type has a certain influence on the algorithm's performance, we adopt GB, KNN (k=3), and SVM algorithms as classifiers. What's more, we compare GADP with GA and other heuristic methods, such as PSO, FPA, and WOA. The hyperparameters and initial settings are described in Table~\ref{parameter}.
We apply the average value and standard deviation to implement a paired two-tailed t-test with other methods or settings. "+", "=" and "-" mean that the objective algorithm performs "better than ", "equal to", and "worse than" the corresponding method When the P-Value is less than 5\%. W/T/L indicates the win/tie/loss statistics of one method relative to another method.

\begin{table}[htbp]
\centering
\caption{The parameter setting of heuristic algorithm}
\renewcommand{\arraystretch}{1.0}
\resizebox{1.0\columnwidth}{!}{
 \begin{tabular}{c c }
	\hline
	Name	&	Parameter Setting		\\
	\hline
   GADP   & max\_feature = 50, initial probability = 0.65, population = 50, max\_iteration = 50\\
   GA     & max\_feature = 50, population = 50, max\_iteration = 50  \\
   PSO    & particles = 50, max\_iteration= 50\\
   FPA    & search\_agents = 50, max\_iteration = 50\\
   WOA    & search\_agents = 50, max\_iteration = 50\\
\hline
	\end{tabular}
}

\label{parameter}
\end{table}

\subsection{The effect of MRMR selection}
\label{5.4}
To evaluate the effect of MRMR, we compare the prediction results of MRMR and the full features based on GADP.
After primary selection, only the top 70 features remain.
In Table~\ref{influence}, we can see that the accuracies of MRMR are higher than the full features on three classifiers except for a few cases.
It denotes that eliminating redundant features is effective and MRMR can select more supportive features to promote the algorithm's performance.
Besides, generally, we regard that a smaller feature scale should have a smaller time cost.
In most cases, MRMR indeed spends less time because the search space of the full features is wider which makes the algorithm converge slower.
As for the full features cases with higher accuracies, we can notice that their time cost usually is higher.
This is because searching for better results needs more time. The phenomenon also appears on the MRMR feature subset.
For example, COIL20 in GB and KNN.

\begin{table}[htbp]
\centering
\renewcommand{\arraystretch}{1.0}
\caption{The influence of MRMR selection}
\label{influence}
\resizebox{1.0\columnwidth}{!}{
	\begin{tabular}{l   l l l l   l l l l    l l  l l    }
	\hline
\multirow{2}*{\textbf{Name}}	&	\multicolumn{4}{c}{\textbf{GB}(\%)}	& \multicolumn{4}{c}{\textbf{KNN}(\%)}	& \multicolumn{4}{c}{\textbf{SVM}(\%)}\\
 \cmidrule(lr){2-5} \cmidrule(lr){6-9}  \cmidrule(lr){10-13}
 &	\multicolumn{2}{c}{\textbf{GADP}}	&  \multicolumn{2}{c}{}  &\multicolumn{2}{c}{\textbf{GADP}}	& \multicolumn{2}{c}{}  &\multicolumn{2}{c}{\textbf{GADP}}\\
 \cmidrule(lr){2-3} \cmidrule(lr){6-7}  \cmidrule(lr){10-11}
&	\textbf{MRMR} & \textbf{full features}	& \textbf{$T_{MRMR}(s)$} 	&	\textbf{$T_{full}(s)$} & \textbf{MRMR} & \textbf{full features}	& \textbf{$T_{MRMR}(s)$} 	&	\textbf{$T_{full}(s)$}&  \textbf{MRMR} & \textbf{full features}	& \textbf{$T_{MRMR}(s)$} 	&	\textbf{$T_{full}(s)$}\\
\hline
PCMAC	&	\textbf{83.53±4.21}	&	61.40±1.60 (+)	&	\textbf{7}	&	39	&	\textbf{87.65±1.31}	&	61.40±1.60 (+)	&	\textbf{234}	&	243 	&	\textbf{89.19±0.95}	&	63.97±1.06 (+)	&	\textbf{74}	&	168 	\\
Isolet	&	\textbf{70.96±1.17}	&	56.67±1.07 (+)	&	\textbf{23}	&	26	&	\textbf{72.18±3.65}	&	56.67±1.07 (+)	&	\textbf{173}	&	274 	&	\textbf{76.47±1.23}	&	62.24±1.63 (+)	&	\textbf{166}	&	612 	\\
GLIOMA	&	\textbf{91.99±6.81}	&	83.97±6.84 (+)	&	\textbf{2}	&	13	&	\textbf{87.98±4.66}	&	83.97±6.84 (+)	&	\textbf{2}	&	12 	&	\textbf{83.97±6.32}	&	78.21±6.70 (+)	&	\textbf{35}	&	176 	\\
Prostate\_GE	&	\textbf{94.19±6.65}	&	65.81±10.77 (+)	&	\textbf{7}	&	12	&	\textbf{96.15±5.44}	&	65.81±10.77 (+)	&	\textbf{7}	&	29 	&	\textbf{93.23±6.55}	&	87.42±10.55 (+)	&	\textbf{32}	&	132 	\\
RELATHE	&	\textbf{82.06±3.04}	&	67.34±1.32 (+)	&	\textbf{9}	&	10	&	\textbf{81.08±1.25}	&	67.34±1.32 (+)	&	\textbf{212}	&	219 	&	\textbf{83.39±1.65}	&	68.68±0.58 (+)	&	\textbf{54}	&	203 	\\
TOX\_171	&	78.99±4.87	&	\textbf{79.58±6.83} (=)	&	\textbf{22}	&	74	&	73.09±4.10	&	\textbf{79.58±6.83} (-)	&	\textbf{33}	&	118 	&	70.17±12.67	&	\textbf{83.67±6.77} (-)	&	\textbf{48}	&	243 	\\
USPS	&	76.61±0.48	&	\textbf{86.57±0.72} (-)	&	\textbf{57}	&	65	&	\textbf{88.95±0.41}	&	86.57±0.72 (+)	&	427 	&	\textbf{131}	&	\textbf{95.92±0.24}	&	90.93±0.61 (+)	&	 \textbf{452}	&	1653	\\
CNAE-9	&	\textbf{88.23±2.42}	&	34.75±1.28 (+)	&	\textbf{9}	&	13	&	\textbf{86.93±0.54}	&	34.75±1.28 (+)	&	150 	&	\textbf{104}	&	\textbf{87.67±1.26}	&	43.56±0.42 (+)	&	\textbf{65} 	&	92	\\
lung\_discrete	&	\textbf{83.48±8.00}	&	80.85±7.04 (+)	&	\textbf{1}	&	2	&	\textbf{97.22±3.21}	&	80.85±7.04 (+)	&	\textbf{1}	&	2 	&	\textbf{94.59±4.30}	&	85.01±6.62 (+)	&	\textbf{65}	&	106 	\\
warpPIE10P	&	\textbf{92.88±5.40}	&	81.40±5.15 (+)	&	\textbf{1}	&	6	&	\textbf{96.65±2.42}	&	81.40±5.15 (+)	&	\textbf{26}	&	74 	&	\textbf{98.57±1.84}	&	 95.71±1.83(+)	&	\textbf{40}	&	42 	\\
lung	&	\textbf{97.55±1.88}	&	94.61±3.71 (+)	&	\textbf{8}	&	16	&	\textbf{97.06±1.96}	&	94.61±3.71 (+)	&	\textbf{53}	&	105 	&	\textbf{96.08±3.20}	&	90.65±1.84 (+)	&	\textbf{75} 	&	103	\\
COIL20	&	\textbf{90.97±1.85}	&	71.74±2.06 (+)	&	59	&	\textbf{37}	&	\textbf{98.54±0.83}	&	71.74±2.06 (+)	&	305 	&	\textbf{276}	&	\textbf{98.96±0.57}	&	96.25±1.12 (+)	&	\textbf{252} 	&	278	\\
warpAR10P	&	\textbf{79.97±11.54}	&	43.09±7.46 (+)	&	\textbf{1}	&	5	&	\textbf{83.83±6.94}	&	43.09±7.46 (+)	&	\textbf{1}	&	8 	&	\textbf{83.81±4.09}	&	75.36±6.86 (+)	&	\textbf{60} 	&	72	\\
Yale	&	\textbf{67.33±7.88}	&	52.15±11.49 (+)	&	\textbf{1}	&	2	&	\textbf{66.71±7.74}	&	52.15±11.49 (+)	&	\textbf{1}	&	4 	&	\textbf{73.39±8.59}	&	70.95±7.45 (+)	&	\textbf{94} 	&	105	\\
Brain\_Tumor\_1	&	\textbf{87.85±6.39}	&	85.62±8.12 (+)	&	\textbf{4}	&	15	&	\textbf{86.66±3.57}	&	85.62±8.12 (=)	&	\textbf{2}	&	10 	&	\textbf{84.44±5.65}	&	83.35±7.39 (=)	&	\textbf{28}	&	132 	\\
W/T/L	&		&	13/1/1	&		&		&		&	13/1/1	&		&		&		&	13/1/1	&		&		\\														
\hline
	\end{tabular}
}
\end{table}

\subsection{The influence of probability setting }

\begin{figure}
 \centering
 \subfloat[0-1\label{one}]{\includegraphics[width=0.3\textwidth]{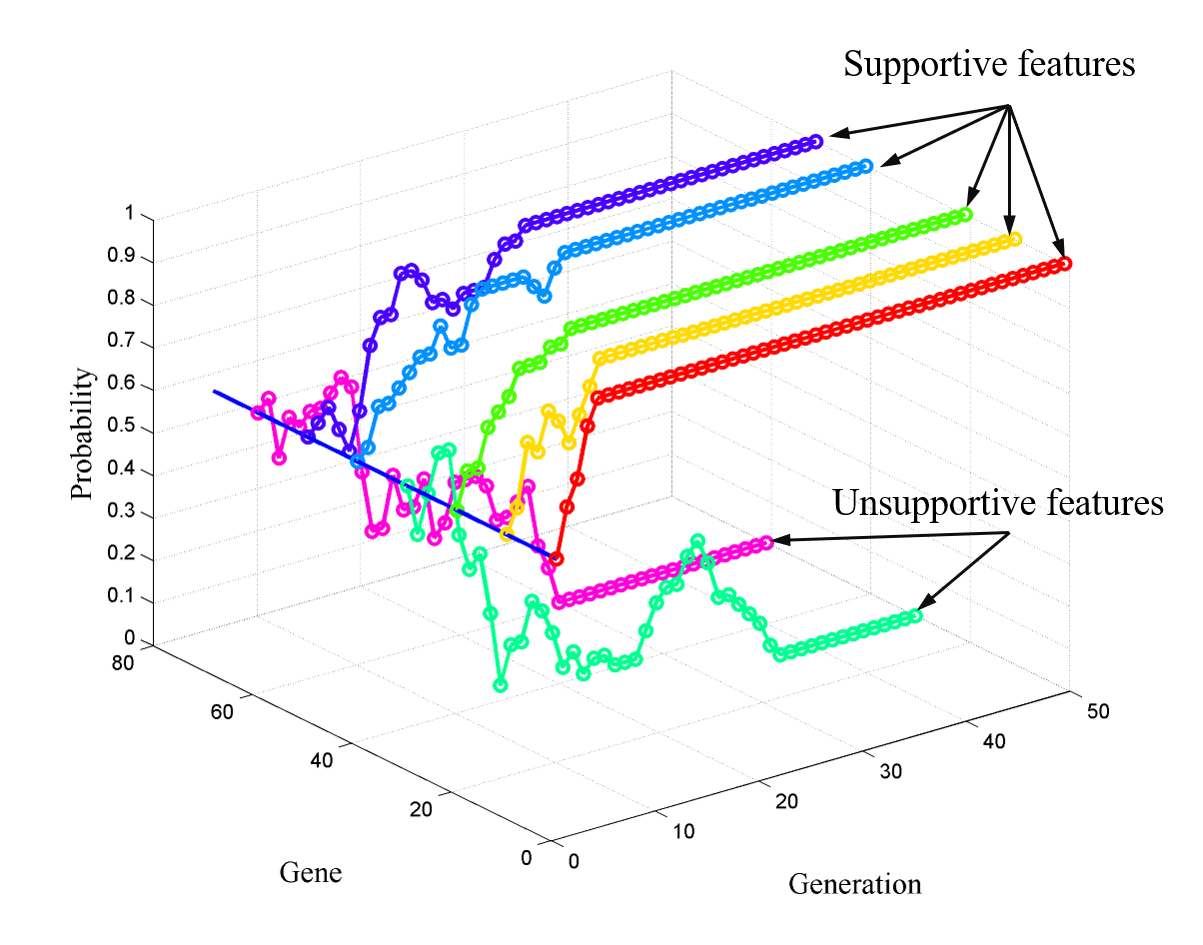}} 
 \subfloat[0.3-0.8\label{two}]{\includegraphics[width=0.3\textwidth]{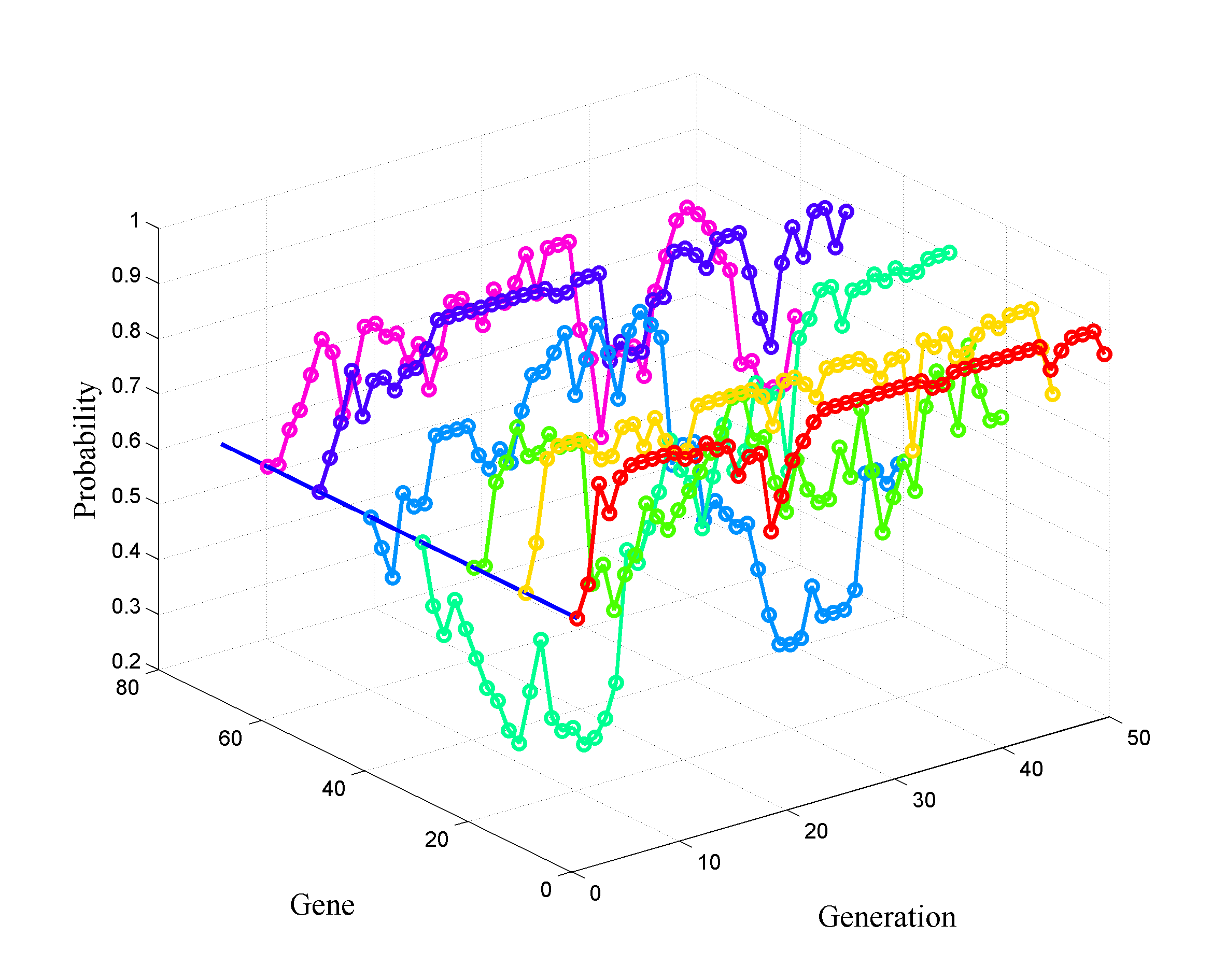}}   
 \subfloat[0.4-0.7\label{three}]{\includegraphics[width=0.3\textwidth]{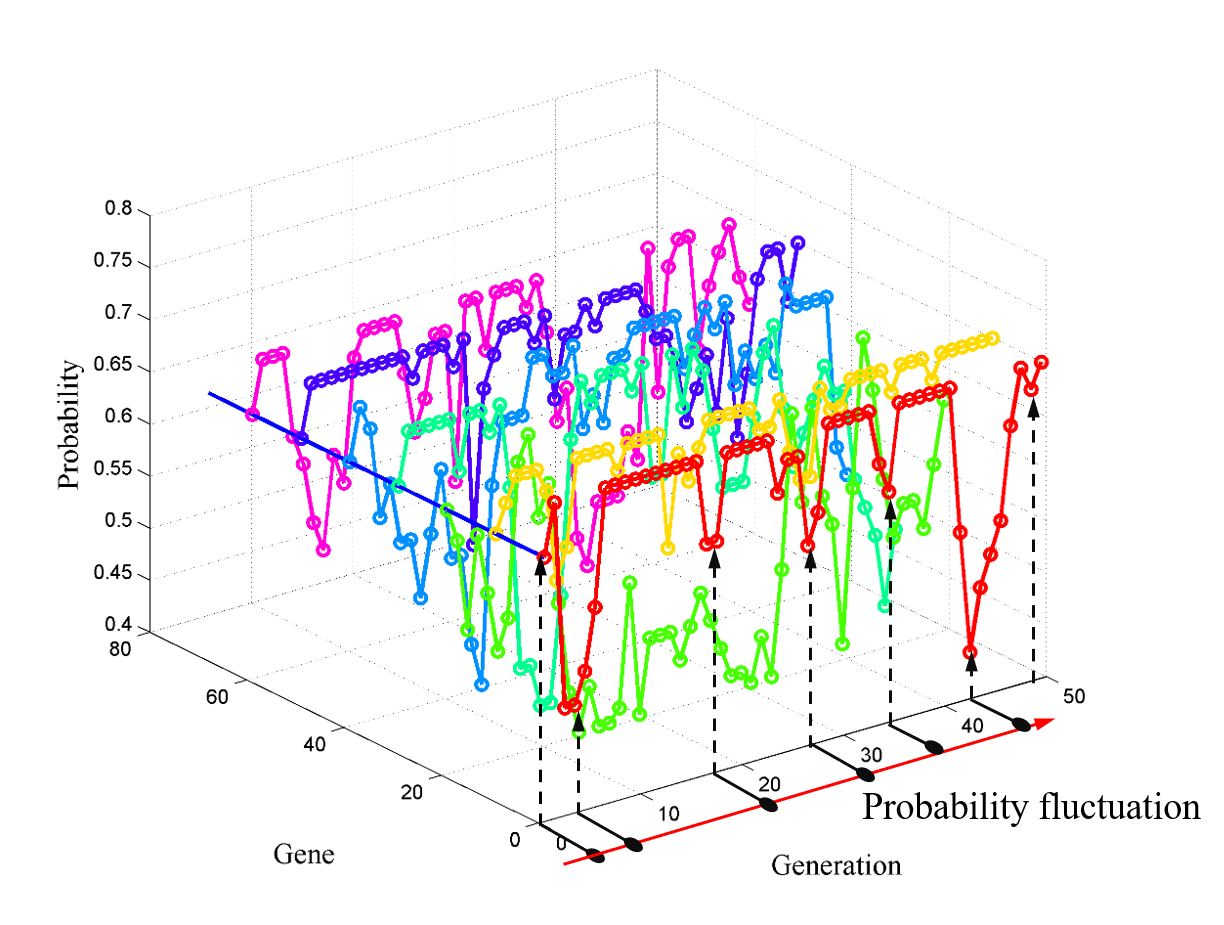}}
 \caption{Gene probability with the increment of generation}\label{probability}
\end{figure}

Gene probability is the core of GADP, which determines the production of each new chromosome.
A proper probability setting is vital for GADP to search the feature space and 3 pairs of probability are studied based on GADP and MRMR in this part.
We depict the tendency of partial chromosomes' probability in Fig.~\ref{probability}. There are three pairs of probability thresholds, 0-1, 0.3-0.8, and 0.4-0.7. We can note that, in the first subfigure, each gene's probability reaches its final value (0-1) after approximately 20 generations. When the selection process is terminated, the results demonstrate that 2 of 7 genes are selected. The earlier convergence of gene probability will decrease the search space, in some way, which could lead to the local optimum.

Compared with the aforementioned figure, the probability fluctuation is obvious in Fig.~\ref{two}, even lasting to the end. The probability could increase or decrease from opposite values.
Finally, we set the gene probability threshold as the 0.4-0.7 pair in Fig.~\ref{three}, which makes converge slower and has a more broad search space, thus, it will cost lots of time. In Fig.~\ref{three}, nearly all the features are under an unstable situation, it seems that the fitness function loses its effect.

Moreover, we also make a comparison of different probability thresholds in Table~\ref{probability_tab}.
The accuracies without the probability limitation are better than those with limitations. They achieve 15, 14 and 10 of 15 with the highest accuracies on GB, KNN, and SVM. It seems that the advantages of 0-1 pair on SVM are not as apparent as on GB and KNN. We think that the SVM has a strong information mining capability which partly decimates the probability threshold's influence.
About the time cost, 0-1 pair spends less average time on GB and SVM. However, on KNN, the situation is different and this should be caused by the influence of KNN's characteristics.
On the whole, we consider that the 0-1 pair's performance is better and adopt it in the rest of the experiments.

%

\begin{table*}[htbp]
\centering   
\renewcommand{\arraystretch}{1.0}
\caption{The influence of gene probability}
\label{probability_tab}
\resizebox{1.0\columnwidth}{!}{
	\begin{tabular}{l   l l l   l l l   l l l    }
	\hline
\multirow{3}*{\textbf{Name}}	&	\multicolumn{3}{c}{\textbf{GB}}	& \multicolumn{3}{c}{\textbf{KNN}}	& \multicolumn{3}{c}{\textbf{SVM}}\\
	 \cmidrule(lr){2-4} \cmidrule(lr){5-7}  \cmidrule(lr){8-10}
\multirow{2}*{}	&	\multicolumn{3}{c}{\textbf{MRMR + GADP}}	& \multicolumn{3}{c}{\textbf{MRMR + GADP}}	& \multicolumn{3}{c}{\textbf{MRMR + GADP}}\\
 \cmidrule(lr){2-4} \cmidrule(lr){5-7}  \cmidrule(lr){8-10}
	&	\textbf{0-1} & \textbf{0.3-0.8}	& \textbf{0.4-0.7} 	&	\textbf{0-1} & \textbf{0.3-0.8}	& \textbf{0.4-0.7} & \textbf{0-1} & \textbf{0.3-0.8}	& \textbf{0.4-0.7}\\
\hline	
PCMAC	&	\textbf{83.53±4.21}	&	82.66±4.09 (+)	&	81.42±4.79 (+)	&	\textbf{87.65±1.31}	&	85.38±1.39 (+)	&	86.67±1.58 (+)	&	\textbf{89.19±0.95}	&	88.57±0.92 (+)	&	88.73±0.71 (+)	\\
Isolet	&	\textbf{70.96±1.17}	&	68.65±1.92 (+)	&	67.56±1.35 (+)	&	\textbf{72.18±3.65}	&	70.71±3.27 (+)	&	70.71±4.23 (+)	&	\textbf{78.01±1.83}	&	76.92±0.89 (+)	&	76.47±1.23 (+)	\\
GLIOMA	&	\textbf{91.99±6.81}	&	88.14±7.93 (+)	&	88.14±7.93 (+)	&	\textbf{87.98±4.66}	&	86.06±3.79 (+)	&	87.98±10.03 (=)	&	\textbf{83.97±6.32}	&	80.98±4.66 (+)	&	82.97±6.32 (=)	\\
Prostate\_GE	&	\textbf{94.19±6.65}	&	\textbf{94.19±6.65} (=) 	&	\textbf{94.19±6.65} (=)	&	\textbf{96.15±5.44}	&	95.15±4.83 (+)	&	94.19±6.65 (+)	&	93.23±6.55	&	\textbf{95.19±7.28} (-)	&	94.15±6.63 (-)	\\
RELATHE	&	\textbf{82.06±3.04}	&	81.71±2.29 (=)	&	81.29±2.32 (=)	&	81.08±1.25	&	\textbf{81.36±0.73} (=)	&	80.87±0.93 (=)	&	83.39±1.65	&	83.53±1.59 (=)	&	\textbf{83.81±2.84} (=)	\\
TOX\_171	&	\textbf{78.99±4.87}	&	71.95±4.86 (+)	&	70.76±2.24 (+)	&	\textbf{73.09±4.10}	&	63.73±1.64 (+)	&	62.54±5.67 (+)	&	\textbf{73.68±4.39}	&	70.17±12.67 (+)	&	67.88±11.93 (+)	\\
USPS	&	\textbf{76.61±0.48}	&	74.10±0.71 (+)	&	73.17±0.52 (+)	&	\textbf{88.95±0.41}	&	88.62±0.41 (=)	&	88.72±0.38 (=)	&	90.93±0.61	&	90.73±0.53 (=)	&	\textbf{90.99±0.53} (=)	\\
CNAE-9	&	\textbf{88.23±2.42}	&	86.75±1.49 (+)	&	85.45±0.97 (+)	&	\textbf{86.93±0.54}	&	85.08±1.32 (+)	&	83.04±1.14 (+)	&	\textbf{87.67±1.26}	&	86.95±1.08 (+)	&	86.93±1.06 (+)	\\
lung\_discrete	&	\textbf{83.48±8.00}	&	83.48±10.25 (=)	&	80.70±9.79 (+)	&	\textbf{97.22±3.21}	&	94.44±6.42 (+)	&	97.22±3.21 (=)	&	85.01±6.62	&	\textbf{91.81±3.05} (-)	&	90.74±3.29 (-)	\\
warpPIE10P	&	\textbf{92.88±5.40}	&	91.46±7.12  (=)	&	91.43±3.93 (=)	&	\textbf{96.65±2.42}	&	\textbf{96.65±2.42} (=)	&	96.18±2.69 (=)	&	\textbf{95.71±1.83}	&	94.75±2.43 (+)	&	95.01±1.83 (=)	\\
lung	&	\textbf{97.55±1.88}	&	\textbf{97.55±1.88} (=)	&	96.08±3.20 (+)	&	\textbf{97.06±1.96}	&	96.56±0.96 (+)	&	96.57±2.47 (+) 	&	\textbf{96.08±3.20}	&	95.09±2.52 (=)	&	94.59±0.94 (+)	\\
COIL20	&	\textbf{90.97±1.85}	&	89.24±1.19 (+)	&	88.61±1.65 (+)	&	\textbf{98.54±0.83}	&	98.13±0.95 (=)	&	97.64±1.03 (+)	&	\textbf{98.96±0.57}	&	97.68±0.76 (+)	&	96.68±0.47 (+)	\\
warpAR10P	&	\textbf{79.97±11.54}	&	79.21±8.60 (=)	&	77.63±13.41 (+)	&	\textbf{83.83±6.94}	&	76.92±5.35 (+)	&	79.14±8.01 (+)	&	\textbf{83.81±4.09}	&	80.75±3.04 (+)	&	82.36±5.63 (=)	\\
Yale	&	\textbf{67.33±7.88}	&	66.72±6.49 (=)	&	64.33±9.62 (+)	&	\textbf{66.71±7.74}	&	63.69±8.28 (+)	&	61.85±10.81 (+)	&	73.39±8.59	&	74.64±10.52 (=)	&	\textbf{77.05±10.65} (-)	\\
Brain\_Tumor\_1	&	\textbf{87.85±6.39}	&	85.62±8.12 (+)	&	84.49±5.57 (+)	&	\textbf{86.66±3.57}	&	85.62±8.12 (=)	&	86.66±3.57 (=)	&	\textbf{83.35±7.39}	&	80.09±7.29 (+)	&	79.00±7.11 (+)	\\
Average Time	&	16.6	&	20.3	&	17.4	&	122	&	80.13	&	70	&	52	&	431.93	&	377.5	\\
W/T/L	&		&	8/7/0	&	12/3/0	&		&	10/5/0	&	9/6/0	&		&	9/4/2	&	7/5/3	\\													
\hline
	\end{tabular}
}
\end{table*}

\subsection{Comparison of GADP with GA and MRMR}
We note that features selected by the MRMR method have a certain redundancy, thus, further selection tries to eliminate the irrelevant features. In this section, we compare the results of GADP algorithm with GA and no process based on MRMR classified by GB, KNN, and SVM.
We depict the average accuracies and standard deviations in Tables \ref{GADP1}-\ref{GADP3}. From accuracy and F1 aspects on three classifiers, the GADP outperforms GA. Both GADP and GA can further promote classifier's performance. Therefore, further feature selection based on MRMR is necessary. About the AUC, MRMR performs better than the results on accuracy and F1. By comparison, we conclude that GADP is superior to the traditional genetic algorithm for feature selection.

\begin{table*}[htbp]
\centering 
\renewcommand{\arraystretch}{1.0}
\caption{Comparsion of GADP with MRMR and MRMR-GA on GB}
\label{GADP1}
\resizebox{1.0\columnwidth}{!}{
	\begin{tabular}{l   l l l   l l l   l l l    }
	\hline
\multirow{3}*{\textbf{Name}}	&	\multicolumn{3}{c}{\textbf{Accuracy}}	& \multicolumn{3}{c}{\textbf{F1}}	& \multicolumn{3}{c}{\textbf{AUC}}\\
\cmidrule(lr){2-4} \cmidrule(lr){5-7}  \cmidrule(lr){8-10}
\multirow{3}*{}	&	\multicolumn{3}{c}{\textbf{MRMR}}	& \multicolumn{3}{c}{\textbf{MRMR}}	& \multicolumn{3}{c}{\textbf{MRMR}}\\
 \cmidrule(lr){2-4} \cmidrule(lr){5-7}  \cmidrule(lr){8-10}
	&	\textbf{GADP} & \textbf{GA}	&  	&	\textbf{GADP} & \textbf{GA}	&  &	\textbf{GADP} & \textbf{GA}	&   \\
\hline
PCMAC	&	\textbf{83.53±4.21}	&	82.66+4.09  (=)	&	77.82±2.22(+)	&	\textbf{87.65+1.31}	&	85.38+1.39 (+)	&	76.85±2.45(+)	&	\textbf{89.19+0.95}	&	88.57+0.92 (+)	&	89.91±6.37(=)	\\
Isolet	&	\textbf{70.96+1.17}	&	68.65+1.92 (+)	&	63.33±1.51(+)	&	\textbf{72.18+3.65}	&	70.71+3.27 (+)	&	61.32±1.51(+)	&	76.47+1.23	&	76.92+0.89 (=)	&	\textbf{97.20±0.28}(-)	\\
GLIOMA	&	\textbf{91.99+6.81}	&	88.14+7.93 (+)	&	75.96±6.38(+)	&	\textbf{87.98+4.66}	&	86.06+3.79 (+)	&	75.36±5.90(+)	&	83.97+6.32	&	80.98+4.66 (+)	&	\textbf{94.62±2.88}(-)	\\
Prostate\_GE	&	\textbf{94.19+6.65}	&	94.19+6.65 (=)	&	92.23±5.37(+)	&	\textbf{96.15+5.44}	&	95.15+4.83 (=)	&	92.20±5.41(+)	&	93.23+6.55	&	\textbf{95.19+7.28} (-)	&	93.18±7.36(=)	\\
RELATHE	&	\textbf{82.06+3.04}	&	81.71+2.29 (=)	&	71.41±0.85(+)	&	81.08+1.25	&	\textbf{81.36+0.73} (=)	&	68.43±1.21(+)	&	83.39+1.65	&	83.53+1.59 (=)	&	\textbf{89.52±0.91}(-)	\\
TOX\_171	&	\textbf{78.99+4.87}	&	71.95+4.86 (+)	&	60.78±4.01(+)	&	\textbf{73.09+4.10}	&	63.73+1.64 (+)	&	59.81±3.29(+)	&	70.17+12.67	&	\textbf{73.68+4.39} (=)	&	\textbf{89.25±2.42}(-)	\\
USPS	&	\textbf{76.61+0.48}	&	74.10+0.71 (+)	&	64.40±1.05(+)	&	\textbf{88.95+0.41}	&	88.62+0.41 (=)	&	64.17±0.85(+)	&	90.93+0.61	&	90.73+0.53 (=)	&	\textbf{92.47±0.47}(-)	\\
CNAE-9	&	\textbf{88.23+2.42}	&	86.75+1.49 (+)	&	85.91±2.61(+)	&	\textbf{86.93+0.54}	&	85.08+1.32 (-)	&	86.02±2.46(=)	&	87.67+1.26	&	86.95+1.08 (+)	&	\textbf{98.23±0.29}	\\
lung\_discrete	&	\textbf{83.48+8.00}	&	83.48+10.25 (=)	&	60.31±9.21(+)	&	\textbf{97.22+3.21}	&	94.44+6.42 (+)	&	51.64±13.23(+)	&	85.01+6.62	&	\textbf{91.81+3.05} (-)	&	68.49±7.17(+)	\\
warpPIE10P	&	\textbf{92.88+5.40}	&	91.46+7.12 (=)	&	89.08±6.56(=)	&	\textbf{96.65+2.42}	&	\textbf{96.65+2.42} (=)	&	88.99±6.94(+)	&	95.71+1.83	&	94.75+2.43 (=)	&	\textbf{97.58±1.62}(-)	\\
lung	&	\textbf{97.55+1.88}	&	97.55+1.88 (=)	&	93.61±2.92(+)	&	\textbf{97.06+1.96}	&	96.56+0.96 (=)	&	93.63±2.72(+)	&	96.08+3.20	&	95.09+2.52 (=)	&	\textbf{97.91±1.95}(=)	\\
COIL20	&	\textbf{90.97+1.85}	&	89.24+1.19 (+)	&	85.21±1.90(+)	&	\textbf{98.54+0.83}	&	98.13+0.95 (=)	&	84.29±2.15(+)	&	\textbf{98.96+0.57}	&	97.68+0.76 (+)	&	98.69±0.25(=)	\\
warpAR10P	&	\textbf{79.97+11.54}	&	79.21+8.60 (=)	&	72.16±12.47(+)	&	\textbf{83.83+6.94}	&	76.92+5.35 (+)	&	71.79±12.25(+)	&	83.81+4.09	&	80.75+3.04 (+)	&	\textbf{94.73±3.12}(-)	\\
Yale	&	\textbf{67.33+7.88}	&	66.72+6.49 (=)	&	58.25±8.45(+)	&	\textbf{66.71+7.74}	&	63.69+8.28 (+)	&	55.25±8.01(+)	&	73.39+8.59	&	74.64+10.52 (=)	&	\textbf{86.65±5.34}(-)	\\
Brain\_Tumor\_1	&	\textbf{87.85+6.39}	&	85.62+8.12 (+)	&	76.78±8.72(+)	&	\textbf{86.66+3.57}	&	85.62+8.12 (=)	&	74.03±12.10(+)	&	\textbf{83.35+7.39}	&	80.09+7.29 (+)	&	81.76±11.02(=)	\\
W/T/L	&		&	7/8/0	&	14/1/0	&		&	7/7/1	&	14/1/0	&		&	6/7/2	&	1/5/9	\\
\hline

	\end{tabular}
}
\end{table*}

\begin{table*}[htbp]
\centering 
\renewcommand{\arraystretch}{1.0}
\caption{Comparsion of GADP with MRMR-GA and MRMR on KNN}
\label{GADP2}
\resizebox{1.0\columnwidth}{!}{
	\begin{tabular}{l   l l l   l l l   l l l    }
	\hline
\multirow{3}*{\textbf{Name}}	&	\multicolumn{3}{c}{\textbf{Accuracy}}	& \multicolumn{3}{c}{\textbf{F1}}	& \multicolumn{3}{c}{\textbf{AUC}}\\
\cmidrule(lr){2-4} \cmidrule(lr){5-7}  \cmidrule(lr){8-10}
\multirow{3}*{}	&	\multicolumn{3}{c}{\textbf{MRMR}}	& \multicolumn{3}{c}{\textbf{MRMR}}	& \multicolumn{3}{c}{\textbf{MRMR}}\\
 \cmidrule(lr){2-4} \cmidrule(lr){5-7}  \cmidrule(lr){8-10}
	&	\textbf{GADP} & \textbf{GA}	&  	&	\textbf{GADP} & \textbf{GA}	&  &	\textbf{GADP} & \textbf{GA}	&   \\
\hline
												
PCMAC	&	87.65±1.31	&	\textbf{87.91±0.99} (=)	&	79.62±1.12(+)	&	87.54±1.32	&	\textbf{87.76±1.02} (=)	&	79.61±1.12(+)	&	\textbf{93.42±1.34}	&	88.27±0.91 (+)	&	86.22±1.63(+)	\\
Isolet	&	\textbf{72.18±3.65}	&	68.85±2.32 (+)	&	68.08±3.11(+)	&	\textbf{71.50±3.65}	&	68.70±2.35 (+)	&	67.57±3.04(+)	&	\textbf{93.53±3.92}	&	92.70±0.88 (=)	&	92.81±0.80(+)	\\
GLIOMA	&	\textbf{87.98±4.66}	&	81.89±7.80 (+)	&	78.04±13.08(+)	&	\textbf{87.57±4.93}	&	81.68±7.69 (+)	&	77.59±8.45(+)	&	92.61±4.56	&	\textbf{95.31±1.96} (-)	&	92.32±4.24(=)	\\
Prostate\_GE	&	\textbf{96.15±5.44}	&	93.19±4.86 (+)	&	90.27±4.87(+)	&	\textbf{96.15±5.45}	&	93.16±4.88 (+)	&	90.22±4.90(+)	&	\textbf{95.78±5.44}	&	95.74±3.82 (=)	&	94.72±5.30(=)	\\
RELATHE	&	\textbf{81.08±1.25}	&	74.63±2.43 (+)	&	79.19±0.91(+)	&	\textbf{81.11±1.26}	&	73.86±2.42 (+)	&	79.23±0.91(+)	&	84.35±1.25	&	76.62±0.82 (+)	&	\textbf{84.59±1.45}(=)	\\
TOX\_171	&	\textbf{73.09±4.10}	&	70.78±6.04 (+)	&	45.60±5.35(+)	&	\textbf{73.05±4.68}	&	70.57±5.29 (+)	&	45.60±5.48(+)	&	87.81±4.49	&	\textbf{88.04±1.93} (=)	&	71.76±5.28	\\
USPS	&	\textbf{88.95±0.41}	&	88.52±0.99 (=)	&	87.97±0.42(+)	&	\textbf{88.87±0.42}	&	88.46±1.00 (=)	&	87.86±0.40(+)	&	\textbf{96.79±0.43}	&	96.74±0.59 (=)	&	96.68±0.22(=)	\\
CNAE-9	&	\textbf{86.93±0.54}	&	84.62±1.41 (+)	&	80.72±2.75(+)	&	\textbf{87.38±0.44}	&	84.80±1.41 (+)	&	81.12±2.90(+)	&	\textbf{95.27±0.51}	&	95.19±1.03 (=)	&	94.21±1.72(=)	\\
lung\_discrete	&	\textbf{97.22±3.21}	&	93.20±2.49 (+)	&	90.35±8.38(+)	&	\textbf{97.14±3.30}	&	92.99±3.75 (+)	&	88.71±11.85(+)	&	\textbf{99.25±3.21}	&	98.03±2.02 (=)	&	98.36±1.61(=)	\\
warpPIE10P	&	\textbf{96.65±2.42}	&	91.42±2.50 (+)	&	94.76±2.39(+)	&	\textbf{96.70±2.38}	&	90.85±2.58 (+)	&	94.81±2.32(+)	&	\textbf{98.44±2.41}	&	98.21±0.64 (=)	&	98.33±1.34(=)	\\
lung	&	\textbf{97.06±1.96}	&	96.06±2.77 (=)	&	94.61±4.35(+)	&	\textbf{96.59±2.28}	&	95.90±2.92 (=)	&	93.69±5.18(=)	&	\textbf{96.23±2.08}	&	96.03±3.61 (=)	&	95.98±3.88(=)	\\
COIL20	&	\textbf{98.54±0.83}	&	97.22±0.51 (+)	&	97.01±0.97(+)	&	\textbf{98.53±0.84}	&	97.20±0.51 (+)	&	96.99±0.98(+)	&	\textbf{99.70±0.83}	&	99.58±0.18 (=)	&	99.62±0.28(=)	\\
warpAR10P	&	\textbf{83.83±6.94}	&	72.28±4.53 (+)	&	70.74±7.47(+)	&	\textbf{82.97±7.29}	&	72.09±4.80 (+)	&	70.25±8.01(+)	&	\textbf{95.92±6.71}	&	94.40±2.76 (=)	&	92.90±3.56(+)	\\
Yale	&	\textbf{66.71±7.74}	&	56.94±9.57 (+)	&	55.17±7.42(+)	&	\textbf{65.69±8.36}	&	55.07±8.84 (+)	&	52.52±7.99(+)	&	\textbf{84.19±8.18}	&	81.68±4.71 (+)	&	82.08±5.15(+)	\\
Brain\_Tumor\_1	&	\textbf{86.66±3.57}	&	85.57±2.00 (=)	&	76.63±6.87(+)	&	\textbf{83.15±3.96}	&	81.17±2.73 (+)	&	70.71±6.76(+)	&	78.52±2.06	&	\textbf{83.03±4.27} (-)	&	78.25±4.07(+)	\\
W/T/L	&		&	11/4/0	&	15/0/0	&		&	12/3/0	&	14/1/0	&		&	3/10/2	&	5/10/0	\\
\hline
	\end{tabular}
}
\end{table*}

\begin{table*}[htbp]
\centering 
\renewcommand{\arraystretch}{1.0}
\caption{Comparsion of GADP with MRMR-GA and MRMR on SVM}
\label{GADP3}
\resizebox{1.0\columnwidth}{!}{
	\begin{tabular}{l   l l l   l l l   l l l    }
	\hline
\multirow{3}*{\textbf{Name}}	&	\multicolumn{3}{c}{\textbf{Accuracy}}	& \multicolumn{3}{c}{\textbf{F1}}	& \multicolumn{3}{c}{\textbf{AUC}}\\
\cmidrule(lr){2-4} \cmidrule(lr){5-7}  \cmidrule(lr){8-10}
\multirow{3}*{}	&	\multicolumn{3}{c}{\textbf{MRMR}}	& \multicolumn{3}{c}{\textbf{MRMR}}	& \multicolumn{3}{c}{\textbf{MRMR}}\\
 \cmidrule(lr){2-4} \cmidrule(lr){5-7}  \cmidrule(lr){8-10}
	&	\textbf{GADP} & \textbf{GA}	&  	&	\textbf{GADP} & \textbf{GA}	&  &	\textbf{GADP} & \textbf{GA}	&   \\
\hline										

PCMAC	&	\textbf{89.19±0.95}	&	87.37±0.43 (+)	&	83.78±0.49(+)	&	\textbf{90.03±1.14}	&	88.48±0.39 (+)	&	84.41±0.77(+)	&	95.91±0.95	&	94.79±0.93(+)	&	\textbf{96.19±0.76}(=)	\\
Isolet	&	76.47±1.23	&	\textbf{76.79±3.26} (=)	&	72.05±0.88(+)	&	76.70±1.23	&	\textbf{77.21±3.05} (=)	&	72.60±1.32(+)	&	98.74±1.51	&	98.60±0.30(=)	&	\textbf{98.75±0.12}(=)	\\
GLIOMA	&	\textbf{85.97±6.32}	&	83.97±4.29 (+)	&	75.13±3.70(+)	&	\textbf{86.19±5.77}	&	84.67±5.03 (+)	&	78.74±5.03(+)	&	93.61±6.17	&	\textbf{95.50±2.60}(=)	&	93.93±3.18(=)	\\
Prostate\_GE	&	93.23±6.55	&	93.15±6.56 (=)	&	87.23±5.37(+)	&	93.77±6.18	&	93.73±6.19 (=)	&	87.85±5.12(+)	&	96.52±6.55	&	96.60±6.80(=)	&	\textbf{97.03±5.52}(=)	\\
RELATHE	&	\textbf{83.39±1.65}	&	80.11±2.84 (+)	&	77.97±1.95(+)	&	\textbf{84.24±1.75}	&	82.34±2.72 (+)	&	78.93±2.34(+)	&	92.19±1.65	&	90.46±1.24(+)	&	\textbf{92.55±1.02}(=)	\\
TOX\_171	&	\textbf{75.17±6.67}	&	64.37±5.53 (+)	&	67.52±6.90(+)	&	\textbf{75.05±6.09}	&	66.15±4.87 (+)	&	67.90±7.57(+)	&	90.34±2.94	&	88.75±2.85(+)	&	\textbf{90.63±2.24}(=)	\\
USPS	&	\textbf{90.93±0.61}	&	88.85±0.57 (+)	&	85.66±0.58(+)	&	\textbf{90.97±0.59}	&	88.88±0.58 (+)	&	85.69±0.59(+)	&	\textbf{99.32±0.62}	&	99.25±0.19(=)	&	99.26±0.10(=)	\\
CNAE-9	&	87.67±1.26	&	\textbf{88.42±2.77} (=)	&	82.95±1.67(+)	&	89.37±1.21	&	\textbf{89.78±2.02} (=)	&	84.19±1.72(+)	&	98.74±1.35	&	98.41±0.50(=)	&	\textbf{98.84±0.26}(=)	\\
lung\_discrete	&	85.01±6.62	&	83.63±4.07 (+)	&	\textbf{85.35±8.38}(=)	&	83.89±8.37	&	79.83±8.37 (+)	&	82.59±13.01(=)	&	98.33±7.62	&	98.63±1.39(=)	&	\textbf{98.68±1.85}(=)	\\
warpPIE10P	&	\textbf{95.71±1.83}	&	94.29±1.54 (+)	&	88.32±3.30(+)	&	\textbf{96.32±1.64}	&	94.42±1.32 (+)	&	89.73±2.47(+)	&	\textbf{99.60±1.83}	&	98.90±0.71(=)	&	99.48±0.55(=)	\\
lung	&	\textbf{96.08±3.20}	&	95.56±3.37 (=)	&	89.11±2.75(+)	&	\textbf{96.17±3.28}	&	95.79±3.09 (=)	&	89.39±2.66(+)	&	\textbf{99.45±3.22}	&	98.80±1.20(=)	&	99.24±1.18(=)	\\
COIL20	&	\textbf{98.96±0.57}	&	98.61±0.39 (=)	&	93.82±0.14(+)	&	\textbf{99.04±0.52}	&	98.71±0.34 (=)	&	93.89±0.14(+)	&	\textbf{99.99±0.57}	&	99.98±0.03(=)	&	99.99±0.01(=)	\\
warpAR10P	&	\textbf{83.81±4.09}	&	82.29±9.91 (=)	&	74.26±3.65(+)	&	\textbf{88.01±2.69}	&	85.08±4.81 (+)	&	76.62±6.13(+)	&	97.15±3.68	&	96.27±3.65(=)	&	\textbf{98.01±1.33}(=)	\\
Yale	&	\textbf{73.39±8.59}	&	70.28±7.90 (+)	&	66.57±9.92(+)	&	76.01±8.29	&	\textbf{76.13±5.61} (=)	&	70.22±12.79(=)	&	90.41±9.07	&	\textbf{92.22±2.01}(=)	&	91.79±3.80(=)	\\
Brain\_Tumor\_1	&	\textbf{83.35±7.39}	&	75.59±1.94 (+)	&	71.73±5.16(+)	&	\textbf{77.09±6.52}	&	65.53±9.03 (+)	&	61.20±10.95(+)	&	\textbf{89.28±6.09}	&	87.28±9.22(=)	&	88.95±4.81(=)	\\
W/T/L	&		&	9/6/0	&	14/1/0	&		&	9/6/0	&	13/2/0	&		&	3/12/0	&	0/15/0	\\
\hline
	\end{tabular}
}
\end{table*}

\subsection{Comparsion of GADP, PSO, FPA and WOA}
In Section \ref{5.4}, we explored the effect of MRMR, the results directly applying GADP on full features are not good. Thus, we compare our method to other heuristic algorithms, such as PSO, FPA, and WOA based on MRMR in Table \ref{heuristic}. On GB and KNN, our proposed method obtains the highest accuracies nearly on all the datasets.  On SVM, GADP achieves 11 of 15 the highest accuracies. Otherwise, it seems that the PSO is a little better than FPA and WOA. What's more, compared with MRMR's results in Table~\ref{GADP1} - \ref{GADP3}, PSO, FPA and WOA also promote their performances as GADP and GA.
Finally, for the feature selection task, we believe that GADP is a considerable method.
\begin{table*}[htbp]
\centering 
\renewcommand{\arraystretch}{1.0}
\caption{Comparsion of GADP, PSO, FPA and WOA}
\label{heuristic}
\resizebox{1.0\columnwidth}{!}{
	\begin{tabular}{l   l l l l   l l l l   l l l l    }
	\hline
\multirow{3}*{\textbf{Name}}	&	\multicolumn{4}{c}{\textbf{GB}}	& \multicolumn{4}{c}{\textbf{KNN}}	& \multicolumn{4}{c}{\textbf{SVM}}\\
 \cmidrule(lr){2-5} \cmidrule(lr){6-9}  \cmidrule(lr){10-13 }
    &	\multicolumn{4}{c}{\textbf{MRMR}}	& \multicolumn{4}{c}{\textbf{MRMR}}	& \multicolumn{4}{c}{\textbf{MRMR}}\\
   \cmidrule(lr){2-5} \cmidrule(lr){6-9}  \cmidrule(lr){10-13 }
	&	\textbf{GADP} & \textbf{PSO}	& \textbf{FPA} 	&	\textbf{WOA}&	\textbf{GADP} & \textbf{PSO}	& \textbf{FPA} 	&	\textbf{WOA}&	\textbf{GADP} & \textbf{PSO}	& \textbf{FPA} &	\textbf{WOA} \\
\hline
PCMAC	&	\textbf{83.53±4.21}	&	81.16±4.23 (+)	&	79.41±3.22 (+)	&	80.24±4.24 (+)	&	\textbf{87.65±1.31}	&	86.36±0.85 (+)	&	81.42±2.62 (+)	&	80.96±1.44 (+)	&	\textbf{89.19±0.95}	&	88.42±0.67 (+)	&	85.95±0.21 (=)	&	87.08±0.30 (=)	\\
Isolet	&	\textbf{70.96±1.17}	&	69.04±2.11 (+)	&	64.87±2.00 (+)	&	65.58±0.97 (+)	&	\textbf{72.18±2.65}	&	70.90±2.74 (+)	&	68.65±3.24 (+)	&	68.40±2.99 (+)	&	\textbf{76.47±0.43}	&	73.28±1.71 (+)	&	72.90±0.55 (+)	&	71.38±1.05 (=)	\\
GLIOMA	&	\textbf{91.99±6.81}	&	90.06±7.72 (=)	&	90.06±7.72 (=)	&	86.06±3.79 (+)	&	\textbf{87.98±4.66}	&	86.06±7.33 (=)	&	83.81±2.09 (+)	&	85.90±1.32 (+)	&	83.97±3.32 	&	\textbf{87.98±4.66} (-)	&	80.13±3.70 (+)	&	81.89±2.63 (+)	\\
Prostate\_GE	&	\textbf{94.19±4.65}	&	\textbf{94.19±4.65} (=)	&	93.23±3.55 (+)	&	93.23±3.55 (+)	&	\textbf{96.15±2.44}	&	95.15±5.44 (=)	&	94.12±2.27 (+)	&	94.15±2.01 (+)	&	93.23±6.55	&	93.23±6.55 (=)	&	94.19±6.65 (=)	&	\textbf{94.23±7.36} (=)	\\
RELATHE	&	\textbf{82.06±3.04}	&	81.50±0.81 (+)	&	77.22±2.43 (+)	&	79.40±2.03 (+)	&	\textbf{81.08±1.25}	&	80.01±3.70 (=)	&	78.77±1.17 (+)	&	77.65±2.13 (+)	&	\textbf{83.39±1.65}	&	80.18±2.20 (+)	&	80.25±1.79 (+)	&	81.57±1.26 (+)	\\
TOX\_171	&	\textbf{78.99±4.87}	&	75.44±2.25 (+)	&	68.41±1.60 (+)	&	69.02±3.32 (+)	&	\textbf{73.09±4.10}	&	61.97±2.50 (+)	&	58.51±4.43 (+)	&	51.44±3.36 (+)	&	\textbf{70.17±2.67}	&	66.16±2.57 (+)	&	67.59±3.31 (+)	&	65.45±6.54 (+)	\\
USPS	&	\textbf{76.61±0.48}	&	75.31±0.53 (+)	&	70.81±1.04 (+)	&	69.51±1.07 (+)	&	\textbf{88.95±0.41}	&	88.44±0.19 (+)	&	88.01±0.68 (+)	&	88.41±0.61 (=)	&	\textbf{90.93±0.61}	&	89.58±0.71 (+)	&	88.62±0.76 (+)	&	88.35±0.51 (+)	\\
CNAE-9	&	\textbf{88.23±2.42}	&	84.25±2.17 (+)	&	79.15±3.64 (+)	&	77.20±2.22 (+)	&	\textbf{86.93±0.54}	&	84.24±1.07 (+)	&	79.43±1.91 (+)	&	78.50±2.65 (+)	&	\textbf{87.67±1.26}	&	83.22±1.09 (+)	&	81.09±1.30 (+)	&	84.43±2.62 (+)	\\
lung\_discrete	&	83.48±8.00	&	\textbf{84.94±9.46} (=)	&	73.76±19.07 (+)	&	75.22±9.88 (+)	&	\textbf{97.22±3.21}	&	94.52±4.54 (+)	&	93.20±2.49 (+)	&	91.89±5.12 (+)	&	87.01±2.62	&	\textbf{90.42±3.46} (-)	&	82.09±4.57 (+)	&	88.96±4.66 (=)	\\
warpPIE10P	&	\textbf{92.88±5.40}	&	91.94±7.26 (=)	&	89.07±5.37 (+)	&	88.61±7.50 (+)	&	\textbf{96.65±2.42}	&	96.18±2.69 (=)	&	96.18±2.69 (=)	&	95.71±3.24 (=)	&	\textbf{95.71±1.83}	&	95.71±3.92 (=)	&	93.79±3.31 (+)	&	94.76±1.39 (+)	\\
lung	&	\textbf{97.55±1.88}	&	95.59±2.35 (+)	&	95.58±1.86 (+)	&	96.08±2.77 (=)	&	\textbf{97.06±1.96}	&	96.57±2.47 (=)	&	96.57±3.35 (=)	&	96.08±3.20 (=)	&	\textbf{96.08±3.20}	&	93.61±2.44 (=)	&	95.09±2.52 (=)	&	95.09±2.52 (=)	\\
COIL20	&	\textbf{90.97±1.85}	&	89.03±2.00 (+)	&	87.08±1.29 (+)	&	87.01±0.62 (+)	&	\textbf{98.54±0.83}	&	97.92±0.70 (+)	&	97.36±0.86 (+)	&	97.01±1.29 (=)	&	\textbf{98.96±0.57}	&	97.89±0.51 (+)	&	97.75±0.66 (+)	&	97.75±0.48 (+)	\\
warpAR10P	&	\textbf{79.97±4.54}	&	77.63±3.41 (+)	&	76.07±4.18 (+)	&	76.87±5.38 (+)	&	\textbf{83.83±6.94}	&	76.16±1.36 (+)	&	73.82±3.40 (+)	&	73.01±9.18 (+)	&	\textbf{83.81±2.09}	&	81.32±2.87 (+)	&	81.53±1.68 (+)	&	81.29±2.08 (+)	\\
Yale	&	\textbf{67.33±4.88}	&	65.51±3.56 (=)	&	62.49±3.02 (+)	&	63.73±4.85 (+)	&	\textbf{66.71±3.74}	&	61.22±3.25 (+)	&	61.85±5.44 (+)	&	58.19±2.11 (+)	&	73.39±8.59	&	\textbf{73.98±7.80} (=)	&	73.37±6.91 (=)	&	71.62±12.93 (=)	\\
Brain\_Tumor\_1	&	\textbf{87.85±2.39}	&	83.35±4.39 (+)	&	82.31±3.28 (+)	&	83.35±2.39 (+)	&	\textbf{86.66±3.57}	&	85.62±8.12 (=)	&	84.54±7.36 (=)	&	82.31±4.68 (+)	&	\textbf{83.35±1.39}	&	81.35±2.39 (+)	&	78.95±3.27 (+)	&	82.26±9.31 (=)	\\
W/T/L	&		&	10/5/0	&	14/1/0	&	14/1/0	&		&	9/6/0	&	12/3/0	&	11/4/0	&		&	9/4/2	&	11/4/0	&	8/7/0	\\
\hline
	\end{tabular}
}
\end{table*}

\section{Conclusion}
\label{Conclusion}

Selecting the optimal subset of features is a vital job for promoting the classifier's performance. Combining the advantages of different methods is a reasonable approach. In this article, we propose a novel population-based algorithm inspired by the genetic algorithm. Different from the traditional genetic algorithm, the crossover and mutation operators are replaced by the dynamic probability. The value of probability assures the existence of heredity and mutation. For evaluating the performance of our proposed method, we explore the influence of probability limitation. Results show that no limation makes the convergence faster and accuracy higher. After, we test the effect of MRMR by comparing GADP's results on the selected features and full features.
The results of MRMR are better than the full features' because there exist lots of irrelevant features which let classifiers trap into local optimum and increase their time consumption.
What's more, we also compare the GADP with GA and no process. Further feature selection can promote the classifier's results. GADP surpasses GA which indicates our proposed method is accessible. Most of cases, the time cost of MRMR is lower than full features', at the same time, with higher accuracy. Sometimes, the results of MRMR have a higher time cost and lower accuracy. This may concern the search space. Larger search space and less supportive features make the classifier's performance bad and cost more time. MRMR filter may lose some key features.
Finally, other famous heuristic algorithms like PSO, FPA, and WOA are compared with our method. Still, our algorithm shows its advantages in feature selection. In these heuristic algorithms, PSO performs better than FPA and WOA.
In the future, we plan to extend our method to other domains.


%
%


\bibliographystyle{spmpsci}      
\bibliography{mybibfile}
\end{document}